\documentclass[a4paper,11pt]{article}
\usepackage[margin=25mm]{geometry}
\usepackage[english]{babel}
\usepackage[utf8]{inputenc}
\usepackage{amsmath}
\usepackage{amsfonts}
\usepackage{graphicx}
\usepackage{latexsym}
\usepackage{graphics}
\usepackage{caption}
\usepackage{cite}
\usepackage{blindtext}

\providecommand{\keywords}[1]
{
  \small	
  \textbf{\textit{Keywords---}} #1
}

\title{\Large\textbf{Mobility Segregation Dynamics and Residual Isolation During Pandemic Interventions}}
\author{Rafiazka Millanida Hilman$^{1}$, Manuel García-Herranz$^{2}$, Vedran Sekara$^{3}$, and Márton Karsai$^{1,4,*}$  \\
        \small $^{1}$Department of Network and Data Science, Central European University, 1100 Vienna, Austria  \\
        \small $^{2}$UNICEF Innovation Office, 10017 New York, USA \\
        \small $^{3}$Department of Computer Science, IT University Copenhagen, 2300 Copenhagen, Denmark \\
        \small $^{4}$National Laboratory for Health Security, Alfréd Rényi Institute of Mathematics, \\
        \small H-1053 Budapest, Hungary \\
        \small Corresponding author: karsaim@ceu.edu
 \\
}
\date{} 

\begin{document}
\maketitle

\begin{abstract}
External shocks embody an unexpected and disruptive impact on the regular life of people. This was the case during the COVID-19 outbreak that rapidly led to changes in the typical mobility patterns in urban areas. In response, people reorganised their daily errands throughout space. However, these changes might not have been the same across socioeconomic classes leading to possibile additional detrimental effects on inequality due to the pandemic. In this paper we study the reorganisation of mobility segregation networks due to external shocks and show that the diversity of visited places in terms of locations and socioeconomic status is affected by the enforcement of mobility restriction during pandemic. We use the case of COVID-19 as a natural experiment in several cities to observe not only the effect of external shocks but also its mid-term consequences and residual effects. We build on anonymised and privacy-preserved mobility data in four cities: Bogota, Jakarta, London, and New York. We couple mobility data with socioeconomic information to capture inequalities in mobility among different socioeconomic groups and see how it changes dynamically before, during, and after different lockdown periods. We find that the first lockdowns induced considerable increases in mobility segregation in each city, while loosening mobility restrictions did not necessarily diminished isolation between different socioeconomic groups, as mobility mixing has not recovered fully to its pre-pandemic level even weeks after the interruption of interventions. Our results suggest that a one fits-all policy does not equally affect the way people adjust their mobility, which calls for socioeconomically informed intervention policies in the future.

\end{abstract} \hspace{10pt}


\section{Introduction}
\label{intro}
\setlength{\parindent}{2em}
\setlength{\parskip}{1em}

Inequality is a prominent feature of today's society. Unequal distribution and access of resources, among others, stand as a preliminary setting. Untangled paths to income ~\cite{luebker2014income}, education ~\cite{smith2014social}, and employment ~\cite{calvo2004effects} seed inequality, which further are moulded into behavioural preferences in daily life,  mostly reflecting proximity to own socioeconomic and demographic background. Eventually, these unequal configurations can lead to segregation that potentially limits the social dynamics.

Socioeconomic segregation is not the only factor that is linked to inequality. There are numbers of ways, such as residential ~\cite{van2020changing}, employment ~\cite{calvo2004effects}, income ~\cite{luebker2014income} or race along which people are segregated, to mention a few. Residential segregation is manifested as separation of different groups of people into different neighbourhoods within a city. Residential segregation is fuelled by the quality of neighbourhoods moving farther away from each other and result in the highly segmented residential places profile between low and high income neighbourhoods ~\cite{musterd2017socioeconomic, tammaru2020relationship}. Therefore, housing plays an intermediary role in reproducing inequality throughout the coupling effects between income inequality and residential segregation ~\cite{tammaru2020relationship}. It has also been shown that growing proportion of high-income segment among workforce increases demand for residential units located in inner city neighbourhoods, due to the centrality of location and accessibility of urban living \cite{butler2008inward, van2020changing}.

Mobility patterns follow on restrictions and preferences on residence and employment in order to meet daily errands. An interplay between inequality and the way people organise their mobility in urban space is inevitable. In line with Urry~\cite{urry2007mobilities}, Olvera et al.~\cite{olvera2004daily} define inequality in mobility as behavioural differences in the level of transport use due to differences in the distribution of monetary ownership such as income or wealth. Furthermore, they find that car ownership is a strong determinant to mobility pattern and residential locations and diminishes potential interaction with people with heterogeneous backgrounds (compared with shared space in public transportation). As a result, segregation patterns come out as an entanglement between inequality and mobility.

In urban mobility network, social stratification in conjunction with unequal access to transport infrastructures brings social exclusion ~\cite{maksim2016mobilities, pooley2016mobility} and social segregation ~\cite{yip2016exploring, mirzaee2020urban}. Such inequalities may change due to external shocks, such as the COVID-19 outbreak, natural catastrophes (earthquakes and floods), or political riots (like war and conflicts). The consequences of such events can dramatically change existing socioeconomic configuration and individual mobility patterns, which in themselves are already constrained by socioeconomic stratification ~\cite{do2021association, chang2021differential, long2021associations}. People's capacities to adjust preferences and their way of living in response to disruptions are limited by their socioeconomic status, limited financial resources or due to their jobs that demand physical presence. As existing literature suggests, people with higher income may have the capacity for larger mobility reduction, while mobility inflexibility and less social distancing are observable among low-income, raising disparity in mobility ~\cite{weill2020social, duenas2021changes, mena2021socioeconomic}. 

In the literature, it is argued that social fabric and inequality shape mobility patterns ~\cite{urry2007mobilities, kaufmann2004motility}. Spatial distribution of commercial areas, residential units, workplaces, and schools, among others, encourages people to move across urban landscape. Built up on the notion of unequal distribution at individual level, mobility is also engendered and reinforced by inequality ~\cite{ohnmacht2016mobilities}. The presence of individual preferences over socioeconomic characteristics of places could be further signified at the socioeconomic (SE) level by taking the visit ratio of people coming from particular SE class to places distributed in various other classes ~\cite{dong2020segregated, hilman2022socioeconomic}. 

We build our approach on this finding by using mobility as an operational concept to analyse socioeconomic stratification and spatial isolation brought by the external shocks. This research investigates the impact of the COVID-19 outbreak and non-pharmaceuticals interventions (NPI) that are later followed in the urban areas of Bogota, Jakarta, London, and New York. Our ultimate goal is to study the changing dynamics of isolation and segregation patterns in mobility due to external shock. We also observe whether such phenomena is temporary, caused by timely restrictions such as lockdown, or they induce long term residual effects. 

To test this, firstly, we capture the changing segregation pattern by quantifying mobility stratification in every sequence of pandemic periods. Secondly, we empirically point out behavioural effects of spatial and socioeconomic exploration in mobility by computing entropy measures derived from spatial and socioeconomic property of visited places. Moreover, we identify types of interventions contributing to aforementioned behavioural effects and their impacts on mobility segregation. Interestingly, these procedures lead us to the still presence of residual effect of shocks even after the removal of interventions.

\section{Results}
\label{sec:res}
\setlength{\parindent}{2em}
\setlength{\parskip}{1em}

In this study we focus on aggregated mobility data that is provided by Cuebiq~\cite{mobility}, a location intelligence and measurement platform (for more details on the data see Materials and Methods). The dataset contains geolocation of places upleveled at census block which were visited by anonymous smartphone users along with timestamps. Time period starts from 1 January 2020 with last day of observation that varies between cities. Given the differences in observation lengths among them,  they all come with the time window that adequately covers an extensive period during pandemics before lockdown, during lockdown, and after reopening as presented in Supplementary Material (SM) Section A. From this dataset, we acquire individual trajectories of 995,000 people with different sample sizes between cities. To detect home location, we use home inference algorithm ~\cite{calabrese2013understanding, csaji2013exploring, phithakkitnukoon2012socio} where home location is defined as the most frequent location visited by each individual during the night time (between 9PM to 6AM). Using this method, we obtain 597,000 of home located people. Consequently, places other than home locations found in the trajectories are classified as place of interest (POI). Details of dataset coverage and the home inference algorithm is specified respectively in Materials and Methods (Section ~\ref{subsec:data} and Section ~\ref{subsec:inference}).

At the same time, we use income related features at spatial resolutions comparable to census tract which are released by respected bureau of statistics, multidimensional poverty index in Bogota ~\cite{mpi2018}, poverty rate in Jakarta ~\cite{bps2018}, total annual income in London ~\cite{ons2015}, and per capita income ~\cite{acs2018} in New York. We combine these mobility data with socioeconomic maps using geospatial information to infer socioeconomic indicator for both people and places. The algorithm pipeline and inference of this study are provided in Materials and Methods (Section ~\ref{subsec:inference}) and SM Section A.

In addition, to quantify policy responses, we use the stringency index released on the Oxford COVID-19 Government Response Tracker (OxCGRT) dataset ~\cite{hale2021global}. Using this data we identify different intervention periods with more or less homogeneous policy restrictions: before lockdown, lockdown, and reopening. 

\subsection{Mobility stratification}
\label{subsec:stratification}

To quantify socioeconomic stratification in mobility, we take the strategy earlier proposed  ~\cite{leo2016socioeconomic, hilman2022socioeconomic} by constructing stratification matrix from mobility network that codes the frequency of visits of people to places. It is defined from their mobility trajectory and indicates the existence of socioeconomic assortativity in visiting patterns. A stratified mobility network is formally constructed as a bipartite structure $G=(U,P,E)$ where individual $u$ is an element of a node set $U$ and place $p$ is constituted to a node set $P$. Visit to $p$ by $u$ is defined as edges $e_{u,p}\in E$, weighted based on the frequency of visit occurrence $w_{u,p}$. In addition, SES of people is defined in terms of the socioeconomic status $c_u=i\in C_U$ of their home location. Following similar method, places are also assigned with a $c_p=j\in C_P$ associated to the socioeconomic status of the census tract of their location.

\subsection{Baseline mobility segregation}
\label{subsec:baseline}
Segregation in the socioeconomic network appears as patterns of assortativity where people of different socioeconomic characters meet less likely than with similar others in the same socioeconomic level. We take the first step to capture stratification tendency by transforming \emph{mobility network} into \emph{mobility stratification matrix} $M_{i,j}$, denoting the probability of people from a given socioeconomic class to visit places with a given socioeconomic class.

As a result, mobility stratification in each period is summarised in a single matrix. To standardise the assortativity measure for the sake of comparability and reproducibility, we compute the mobility assortativity index $r$ defined as a correlation coefficient of $M_{i,j}$ ~\cite{newman2003mixing, dong2020segregated, bokanyi2021universal}. Assortativity index values closer to one signal the higher concentration of visiting venues closer to one's own socioeconomic range (assortative mobility), while 0 pinpoints the dispersion in visiting pattern throughout classes (non-assortative mobility). Otherwise, negative values indicate the tendency to visit places opposite one's own socioeconomic class (disassortative mobility). Complete technical note on transformation technique and assortativity computation is discussed in Materials and Methods (Section ~\ref{subsec:matrix}).

\begin{figure*}[ht!]
    \centering
        \includegraphics[width=0.9\linewidth]{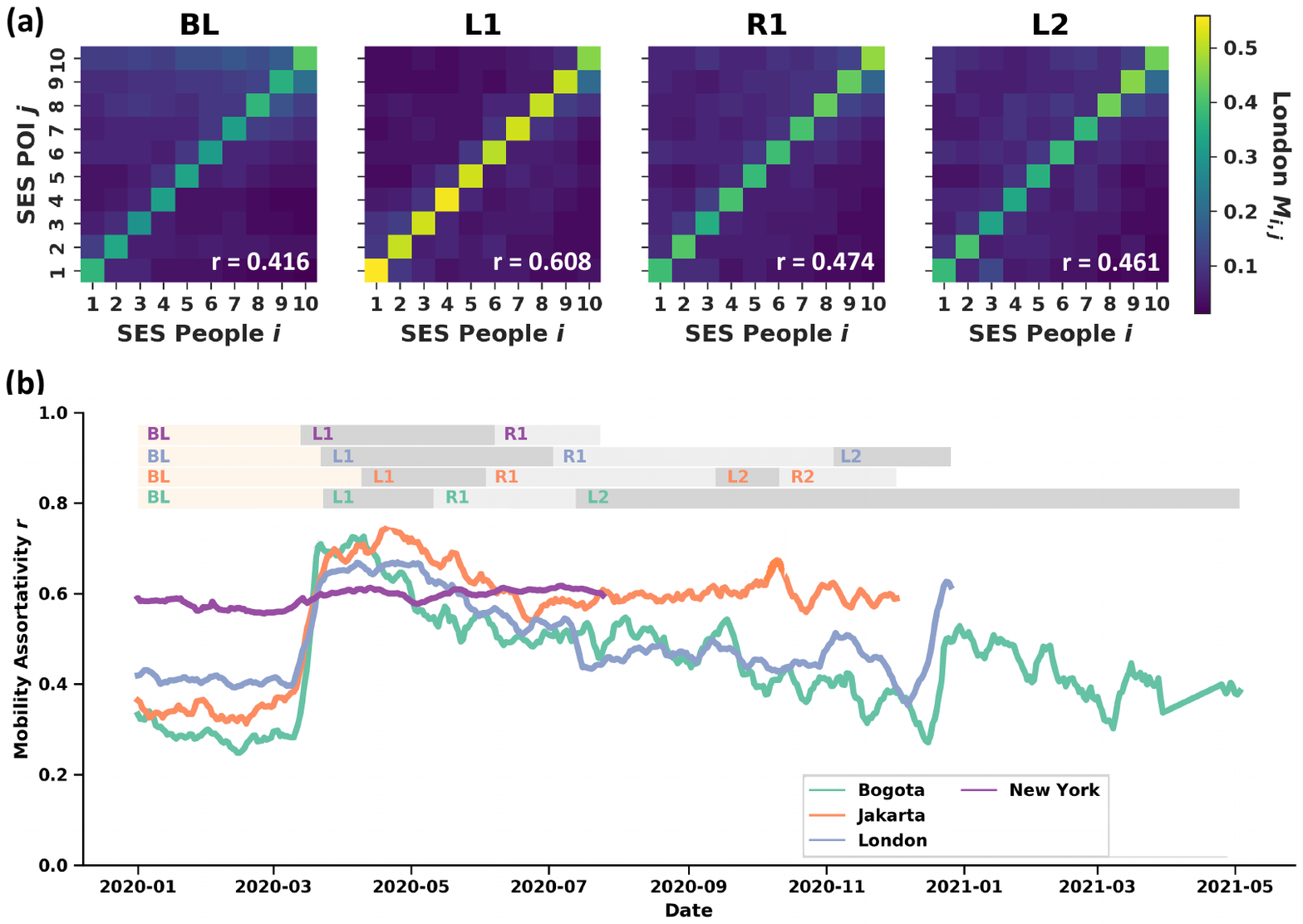}
        \caption{\footnotesize\textbf{Mobility stratification matrix $M_{i,j}$}. The structure of empirical socioeconomic stratification in London is visualised in a matrix form composing visit probabilities of individuals in each class to places located in various other classes. Fig.~\ref{fig:1}a reveals that larger visit proportion happens in a bin with lighter colour grades along diagonal elements across periods: Before Lockdown (BL), Lockdown (L1/L2), and Reopening (R). The strength of assortative mixing is quantified by a correlation coefficient between $i$ and $j$ denoted as $r$. We find stronger diagonal concentration during lockdown, denoting considerable visits to locations within own SES. Therefore, enforcing lockdown levels up assortative mixing. This is considered as a change in mobility preference due to NPI. Fig.~\ref{fig:1}b is constructed by implementing sliding window algorithm. For every 1 week window with 1 day slide interval, a mobility matrix is generated with computed $r$. Increasing $r$ overlaps with lockdown period. Colour shades of line and block denotes city. }
        \label{fig:1}
\end{figure*}

To demonstrate these metrics and to follow up on the dynamical changes of segregation during different phases of crisis interventions, we take the example of London. Fig.~\ref{fig:1}a provides snapshots of mobility stratification patterns in London, starting from before lockdown and followed by the interchangeable periods between series of lockdown and reopening. In Fig.~\ref{fig:1}a, x-axis represents socioeconomic classes of people $i$ while y-axis denotes socioeconomic classes of places $j$ they visited. As people move, we calculate the frequency of visits for each pair of classes (people-place), proportional to total visits made by everyone who belongs to $c_u=i$ (column-wise normalisation). Colour shades differ the visit magnitude where it becomes lighter as visit proportion gets larger. Fig.~\ref{fig:1}a contains all locations in the trajectories, regardless being home or non-home areas. 

Note that to refine the observation, we isolate home location effects on visiting pattern by removing own home location from mobility trajectory of each individual. The computational result of this sanity check shows weaker but consistent segregation pattern (see SM Section B.2). Assortative mixing is consistently pronounced regardless types of policy imposed on mobility restriction, for instance lockdown and reopening. Moreover, it validates the finding as the revolving pattern persists even after we exclude own home location from mobility trajectory of each individual. 

We consider next the persistence of the segregation patterns during the baseline period. Here we use the baseline segregation level shown by the mobility assortativity $r$ value during Before Lockdown (BL) as the reference point to which the changing patterns in segregation could be adequately compared. Looking at the first matrix in Fig.~\ref{fig:1}a, we obtain an assortativity index $r$ = 0.416, indicating baseline segregation in mobility where to be fairly large, due to the visits that are concentrated on areas with similar SE status to of the visitors', even they were far from their home location.

Subsequently, we continuously observe how segregation changed daily over an extensive period before the COVID-19 pandemic. In Fig.~\ref{fig:1}b, we look at more granular temporal length by using sliding windows to construct a sequence of daily mobility stratification matrix (Fig.~\ref{fig:1}b). For every 2 weeks window with 1 day slide interval, we create a matrix and measure its assortativity index $r$. The initial value of $r$ indicates respectively Bogota (green), Jakarta (orange), London (light blue), and New York (purple). Looking at the baseline assortativity index values, New York stands out with $r$ around 0.571, while Bogota reaches $r$ value around 0.317. Assortativity degree in daily individual mobility in Jakarta is about 0.366 and London records the $r$ value approximately at 0.416. Apart form that, we see that the assortativity level in mobility during baseline period tends to be constant without remarkable jump or drop between days. 

\subsection{Segregation dynamics due to external shocks}
\label{subsec:changing}

As we can see on Fig.~\ref{fig:1}b, the assortativity index $r$ sensitively reflects changes in mobility segregation during different intervention periods. More prominently, the implementation of lockdown (L1 and L2), harnessed mobility at large and encouraged people to visit POI within their own socioeconomic spectrum. This leads the coefficient $r$ to reach its peak at 0.608 during the first lockdown (L1) in London, after a 46\% increase from its baseline level at 0.416. In this city,  mobility is reintroduced during reopening (R1), and visiting more places became possible again. Chance for higher socioeconomic mixing in mobility was opened, resulting in lower $r$ at 0.474. However, it has not retrieved back to the original level before lockdown but remaining 14\% higher than the baseline level. A weaker impact of lockdown were found during the second phase (L2) even resulting in a in 11\% higher $r$ at 0.461 as compared to the baseline level. We recognise this phenomena as induced assortativity. Similar matrices computed for other urban areas are presented in SM Section A.1. 

The general overview of assortativity dynamics in Fig.~\ref{fig:1}b indicates that mobility assortativity is found in all investigated cities except New York. Since the implementation of lockdown policy onward, increase in $r$ value in Bogota was visible with the highest value recorded at 0.613 during the first phase of lockdown. It suggests that the large spike of visitation to places located in own socioeconomic status.  In the following periods, $r$ value tended to stabilised around 0.5, still higher than the baseline level. In Jakarta, once the lockdown was introduced, $r$ value was staggering around 0.6 in the periods that came after. The intermittent reopening phase only decreased the $r$ value temporarily and it surged again after the second phase of lockdown was taken into account. In the end, the $r$ value was still twice larger than the original magnitude before lockdown. 

Mobility assortativity in New York remained relatively stable across the time without any significant temporal cycle. This invariant pattern in New York could be accounted to the imbalance and asymmetric mobility between five boroughs within its territory: Manhattan, Brooklyn, Queens, Bronx, and Staten Island. In related studies, Rajput and other~\cite{rajput2021revealing} state that stay-at-home orders implemented in the midst of COVID-19 outbreak disturbed 80\% typical daily movement within city in New York from as early as the second half of March 2020. Recalling that Manhattan is the epicentre of the city's human dynamics where various mobility motifs and activities occur, we observe the case on Manhattan separately along with mobilities within and between other boroughs in New York. Our results are summarised in the SM Section E to clarify the upsurge in assortativity during lockdown that already found in other cities.

\subsection{Residual isolation}
\label{subsec:residual}

\begin{figure*}[ht!]
        \includegraphics[width=0.9\linewidth]{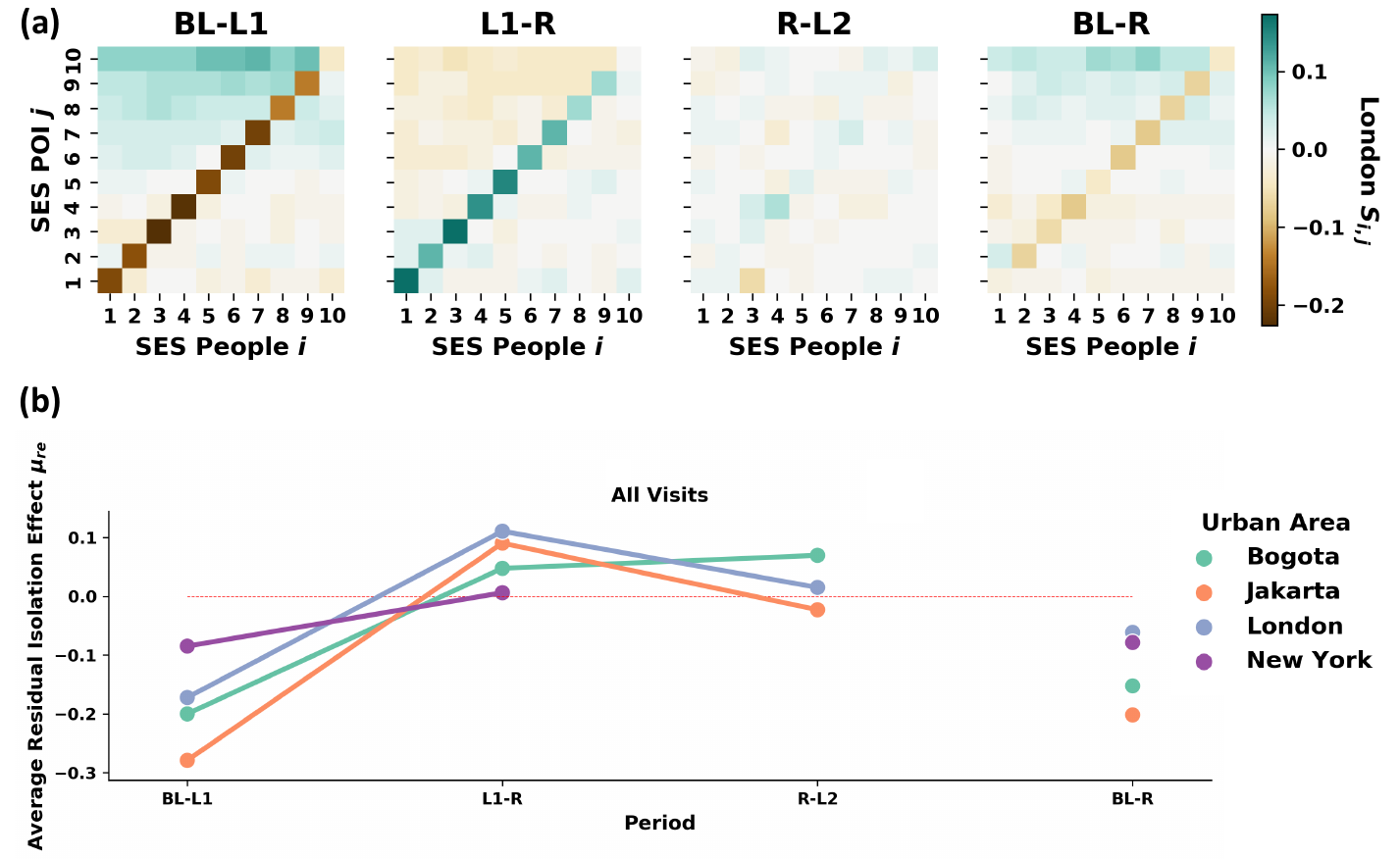}
        \caption{\footnotesize\textbf{Mobility adjustment matrix $S_{i,j}$}. It shows the ratio in stratified mobility pattern between a period during the pandemic namely Lockdown (L1/L2) and Reopening (R) as to compare to  Before Lockdown (BL). Green shades indicate more visits  made before the enforcement of lockdown, white blocks constitute equal visits, otherwise brown blocks appear. Therefore, we observe contrast proportion on the upper diagonal elements in London as visits to these places touch the lowest level in L1 relative to BL, burst in R1 and drop in L2 (Fig.~\ref{fig:2}a). Residual isolation effects as measured by average value of main diagonal trace in each matrix $\mu_{re}$. Comparative measure across cities in terms of average residual isolation effect $\mu_{re}$ is provided in Fig.~\ref{fig:2}b. Purple block shows the difference between before lockdown baseline and reopening stage.}
        \label{fig:2} 
\end{figure*}

To further refine the observation related to changing segregation pattern, we measure the presence of \emph{residual isolation}. The ultimate recovery is expected when mobility pattern and assortative mixing during the reopening stage are on the same level as before lockdown. If such conditions hold, sudden changes triggered by external shock namely COVID-19 outbreak might only carry short-temporal effect inducing any barrier for people to return to the normal pre-pandemics configuration. To quantify such effects we define the \emph{mobility adjustment matrix} $S_{i,j} = M_{i,j}^{t1} - M_{i,j}^{t2}$ is set by taking the difference between \emph{mobility stratification matrix} $M_{i,j}$ in two consecutive periods, for instance between baseline period $M_{i,j}^{BL}$ and the first lockdown $M_{i,j}^{L1}$. Therefore, the matrix element $a_{ij}$ in $M_{i,j}$ entails the difference in proportion of frequency of visits between a pair of consecutive periods as seen in Fig.~\ref{fig:2}.

Fig.~\ref{fig:2}a reveals the difference between a pair of intervention periods before lockdown and the first lockdown, inferring that the first lockdown is the most stringent among others. It tells us that the induced assortativity develops into isolation. In case of London, the upper diagonal elements of $S_{i,j}$ are dominated by negative value, indicating as away less visits to these places located in the higher socioeconomic class during the first lockdown as compared to the baseline level. The arrival of the second lockdown period pushes the visiting proportion to higher SES places to a lower level again, but not as large as in the first lockdown. The relaxation on mobility restriction during the reopening period increases the visits to these places to an extent, although negative values are still found in some cells.  

Quantitative measure of residual isolation $\mu_{re} = \frac{\sum tr[M_{i,j}]}{\sum_{j\in C_P}}$ is provided by taking the summation over main diagonal elements of $M_{i,j}$ and divide the value by the number of socioeconomic classes which is ten. It results in the average value of matrix diagonal elements as shown in Fig.~\ref{fig:2}b. 

In each city even in New York, in the extreme degree, individuals during lockdown restrict their preference to be present in the areas within own socioeconomic boundary more than they used to be. As the reopening is imposed after the first lockdown, the pattern is reversed. The difference between reopening and the second lockdown is very subtle. Interestingly, the reopening is not necessarily able to restore the typical configuration to before lockdown. We still see negative value along main diagonal traces, even higher than -0.2, as shown by the negative diagonal gradient, revealing the existence of residual isolation effect. In Jakarta, people tend to spend almost more than 30\% frequent activities in the class they belong to. Average residual effect in Bogota, is captured around 20\% and nearly about 10\% in New York. However, the reopening (compared to before lockdown/BL-R) does not directly bring $\mu_{re}$ equal to zero in any cities we observe, indicating the prevalent residual isolation. Weaker average residual isolation is found after removing local visits (see SM Section B.2) and pushes $\mu_{re}$ closely distributed around zero.

\subsection{Restriction and behavioural effects}
\label{subsec:restriction}
Pandemic brings another complexity in the way people move from one location to numerous others across space. During the COVID-19 outbreak, mobility is not merely driven by established personal preference but also supplementary necessity to align with prescribed mobility restrictions. 

\begin{figure*}[ht!]
    \centering
        \includegraphics[width=0.9\linewidth]{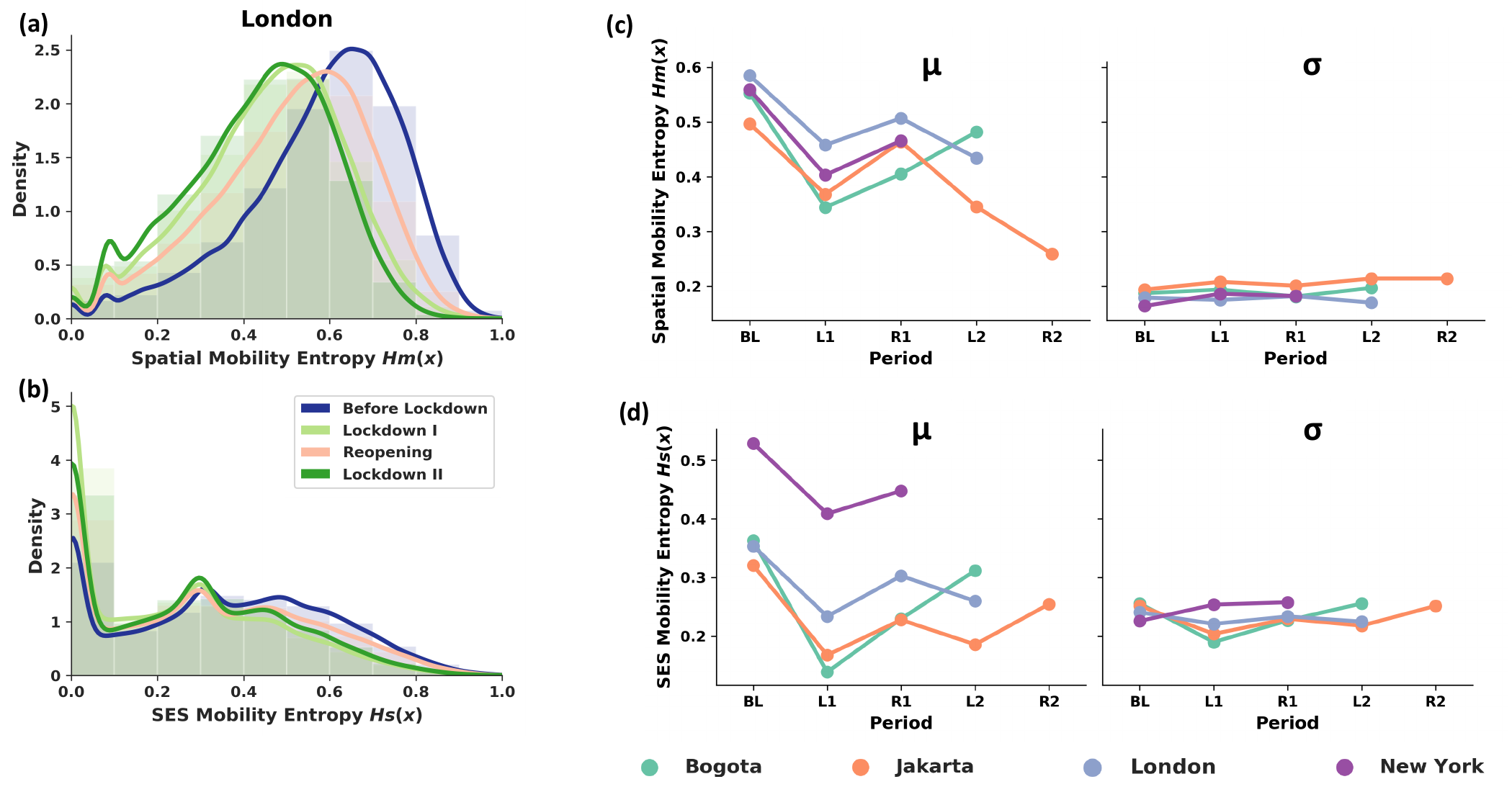}
        \caption{\footnotesize\textbf{Spatial and SES mobility entropy.} Spatial mobility entropy ${H_m(X)}$ (Fig.~\ref{fig:3}a) takes into account the heterogeneity of places in individual trajectory with value range from 0 (visiting same locations) to 1 (visiting various locations). SES mobility entropy ${H_s(X)}$ (Fig.~\ref{fig:3}b)  takes similar computation after replacing set of locations with socioeconomic status of area where those places located implying visit variation between socioeconomic isolation (0) and socioeconomic diversity (1). In London, we observe less heterogeneity in both locations and socioeconomic status of places visited by individual during lockdown. Even after some relaxations are allowed, people do not experience mobility at pre-pandemics level. Similar observation also become evident in other cities globally  (Fig.~\ref{fig:3}c).}
        \label{fig:3}
\end{figure*}

With this in mind, we look at heterogeneities of where-to-go decision from two different aspects: spatial and socioeconomic composition. We use an entropy based measure, which we develop on top of Shannon's formula, to measure the heterogeneity of mobility traces in term of geolocation. Here we define  \emph{spatial mobility entropy} ${H_m(X) = - \sum_{x \in X} p_{(x)} \log_2 p_{(x)}}$ where geolocation and SES is $x \in X$ and \emph{SES mobility entropy} ${H_s(X) = - \sum_{x \in X} p_{(x)} \log_2 p_{(x)}}$ for which socioeconomic class is $x \in X$. In the formalisation of \emph{spatial mobility entropy} ${H_m(X)}$, we compose a scalar for each individual trajectory containing geographic location of places visited a single people. For \emph{SES mobility entropy}, we replace the geographic location information with socioeconomic classes where visited places belong to. In both types of entropy, lower values correspond to higher domination of particular locations/SES of locations in the visit pattern, signalling the extensive locational/socioeconomic isolation. Given that the measure is normalised by period, the upper cut-off is 1 (absolute heterogeneity) and the lower cut-off is 0 (absolute homogeneity). Formal formulation of entropy is available in Materials and Methods (Section ~\ref{subsec:entropy}).

As shown in Fig.~\ref{fig:3}a and b, in London, we deal with four phases of pandemic: Before Lockdown (BL), Lockdown I (L1), Reopening (R1), and Lockdown II (L2). While Fig.~\ref{fig:3}a reveals the distribution of locational mixing degree in individual trajectory. Fig.~\ref{fig:3}b follows the similar way but rather emphasising on socioeconomic setting of those listed locations. In both figures, skewness of the curve moves to the left (to the direction of zero) in the first lockdown (light green), so does in the second lockdown (dark green). It points out the tendency of upholding more homogeneous visiting pattern. In respect of spatial scale, urban explorability drops once policy limiting mobility flow implemented. Consequently, the set of visited places becomes more narrow (centred to smaller set of places) and localised (closer to where home is located). Similar pattern also holds with regard to socioeconomic range. As set of locations is shrunken by distance, it becomes highly concentrated to particular socioeconomic level that reflects own well-being.

We check the shifting magnitude by computing the average value ($\mu$) and standard deviation ($\sigma$) of the two entropy distributions for the different cities. In Fig.~\ref{fig:3}c, the initial phase of lockdown (L1) characterises mobility pattern to be locationally more homogeneous since spatial mobility entropy ${H_m(X)}$ is lower than before lockdown period (BL). Spatial concentration largely happened in Bogota during L1, reaching the average value at 0.35. Jakarta recorded the average spatial diversity at 0.37. In addition, the average value in New York and London respectively was around 0.4 and 0.5. The reopening phase that follows (R1) does not bounce the variability of locational and socioeconomic preference back to original level before lockdown even though it goes to recovery direction. Compared to spatial mobility entropy, SES mobility entropy ${H_s(X)}$ in Fig.~\ref{fig:3}d receives grave repercussions caused by the outbreak even more as $\mu$ ranges from about 0.5 to lower values. During L1, People in Bogota and Jakarta experience deeper socioeconomic isolation as ${H_s(X)}$ falls below 0.2. London is close to 0.35 while New York is around 0.4.  

\subsection{Mobility interventions}
\label{subsec:intervention}

\begin{figure*}[ht!]
    \centering
        \includegraphics[width=0.9\linewidth]{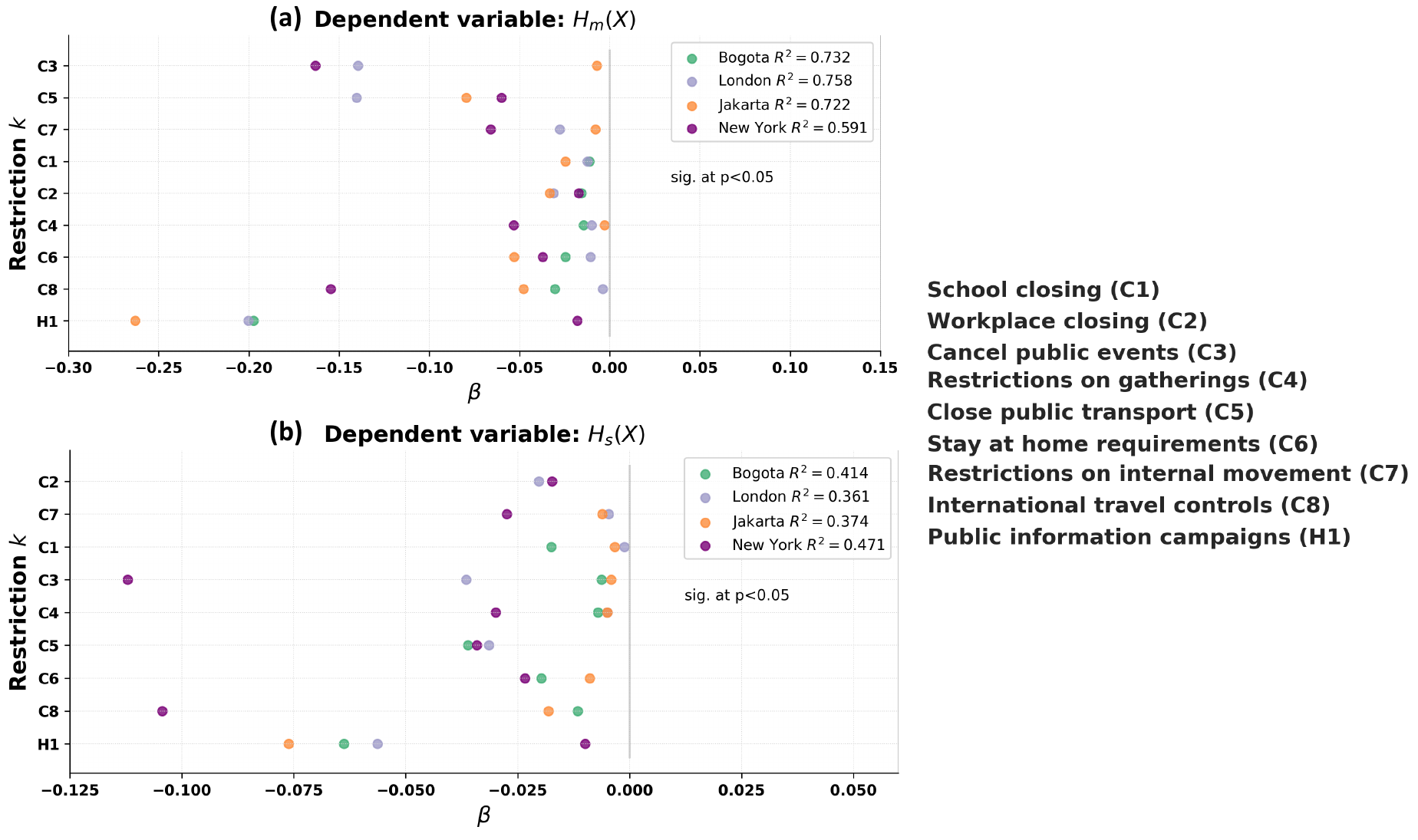}
        \caption{\footnotesize\textbf{Multivariate regression.} The effectiveness of NPI in constraining spatial exploration ${H_m(X)}$ (Fig.~\ref{fig:4}a) and socioeconomic exploration ${H_s(X)}$ (Fig.~\ref{fig:4}b) is presented as covariates $\beta$. In all cities except New York, public information campaign (HI/light purple) is the most influential instrument that highly affect both spatial and economic exploration. The $R^2$ of respected regression models namely for ${H_m(X)}$ and ${H_s(X)}$ differs across cities. The nine types of restrictions explain around 59\% to 76\% of the variance in spatial exploration and turns out lower in socioeconomic exploration from 36\% to 47\%.}
        \label{fig:4}       
\end{figure*}

To this point, we have revealed residual isolation effects of shocks even after mobility restrictions were gradually lifted. However, the kind of restriction that significantly contributes to such configuration is still unknown. Data on NPI ~\cite{hale2021global} contains the strictness level of every single restriction $k$ = 9 categories over period of time, including closing of main venues such as school, workplace, and others. For a complete list see Table 1 in SM Section A. We weight the impact of those restrictions listed as NPI by running multivariate linear regression where the dependent variable an entropy (${H_m(X)}$ or ${H_s(X)}$) and the independent variable a stringency level of each restriction $s_k \in S_K$. The methodological definition for this approach is further explained in Materials and Methods (Section ~\ref{subsec:ranking}).

Individual exploration occurs not only over socioeconomic dimension, but also beyond physical space. Therefore, enforcement of mobility restrictions NPIs also reduce socioeconomic diversity of visiting places. Indeed, from the results shown in Fig.~\ref{fig:4}, public information campaign (H1/light purple) is the most preponderant in each city, simultaneously affecting mobility in terms of spatial and socioeconomic diversity of visited places. 

However, the magnitude that public information campaign restriction brings to mobility is not uniform between physical and socioeconomic space. The covariates ratio is defined as $\beta_{m,s} = \frac{\beta^k_{H_{m(X)}}}{\beta^k_{H_{s(X)}}}$ to indicate relative impact of a type of restriction on those two aspects of exploration. Once this restriction is imposed in London, for instance, its impact on the shrinking spatial diversity in individual trajectory is 3.33 times higher. This number is 3.08 in between Bogota and 3.47 in Jakarta. Meanwhile in New York, the cancellation of public events (C3) concurrently diminishes spatial exploration 1.33 times more than socioeconomic exploration. 

Looking at the $R^2$, we find that the overall values are lower for the model with dependent variable of SES mobility entropy ${H_s(X)}$ as compared to the one fitting on spatial mobility entropy ${H_m(X)}$. We compute ratio values  of $R^2$ for ${H_m(X)}$ over $R^2$ for ${H_s(X)}$, formally expressed as $R^2_{m,s} = \frac{R^2_{H_{m(X)}}}{R^2_{H_{s(X)}}}$. 

In Bogota, the same set of NPI explains a much higher variance of ${H_m(X)}$, 1.76 more than the variance of ${H_s(X)}$. Similar range of ratio values of $R^2_{m,s}$ is also obtained in London (2.10), Jakarta (1.93), and New York (1.25). As the results show that composition of socioeconomic preference over places in individual visiting patterns is still largely shaped by unobserved factors other than mobility restriction, it could be an indication that socioeconomic exploration incorporates more complex dimension than the delineation spatial boundary alone. 

\section{Discussion and conclusions}
\label{sec:disscussion}
\setlength{\parindent}{2em}
\setlength{\parskip}{1em}

In this study, we took a step to analyse the impact of COVID-19 outbreak on structural preference reflected in mobility pattern by looking at the mobility dynamics in Bogota, Jakarta, London, and New York. We found that in-class visits dominate mobility pattern in every temporal snapshots, ranging from before lockdown, lockdown, to reopening. Dependency patterns of assortative behaviour dependencies were also detected as the assortativity coefficient $r$ remained highest during lockdown. Subsequently, the emergence of reopening did not directly bring the typical mobility mixing pattern to the original level observed before the enforcement of lockdown, indicating the existence of residual isolation effect. 

We further measured the degree of residual isolation by comparing stratification in mobility pattern between two consecutive periods (see Fig.~\ref{fig:2}a). It validated the presence of residual isolation effects where visits within own class during reopening is still higher than the usual rate. Another feature of isolation in mobility that has been presented in this study is the decreasing heterogeneity of where-to-go decision from two distinctive aspects: spatial and socioeconomic composition (see Fig.~\ref{fig:3}). Entropy measures revealed that visits became highly concentrated to particular locations and socioeconomic classes. 

To understand which type of NPI does constrain mobility across time window, we proposed multivariate regression model composing all mobility restrictions to examine their magnitudes in intervening the diversity configuration of visiting patterns. In cities, except New York, we observed the impact of public information campaign (H1) gained its highest importance among any other type of restrictions. The observed variability of magnitude could be related to the structure of urban fabric in respected city as well as the level of socioeconomic well-being.

Apart from the computations demonstrated to this point, we realise that stronger evidence for residual isolation in the longer term could be presented if the access to more recent data is available. Our latest data only covers the initial period of reopening where NPI and the COVID-19 protocols were still at the frontier in controlling the outbreak. It solely depends on the behavioural conformity and attitude towards mask wearing and social distancing without any intervention from vaccination policy. Another boundary that we would like to underline is the limitation in direct comparison between cities. This issue is raised due to the different metrics and levels of spatial resolution we use to define SES indicators, that are strongly depending on the availability of data. 

This study contributes to the scientific importance in refining the impact of pandemic on the reorganisation of mobility segregation. It allowed us to comprehensively understand potential occurrence of residual isolation during pandemic interventions at higher spatial and temporal resolution. Afterwards, it taps the pivotal aspect of societal impact as additional detrimental effects induced by residual isolation might not be equally distributed across socioeconomic class, indicating a higher vulnerability faced by lower socioeconomic class that should be better mitigated by adaptive policy design in the future. Therefore, as a future goal, we consider the importance of conducting class-wise analysis to study how different classes are impacted differently. 

\section{Materials and methods}
\label{sec:materialsandmethods}

\subsection{Data description}
\label{subsec:data}

Mobility data is provided by Cuebiq, a location intelligence and measurement platform. Data were shared under a strict contract with Cuebiq through their Data for Good COVID-19 Collaborative program where they provide access to de-identified and privacy-enhanced mobility data for academic research and humanitarian initiatives only. Mobility data are derived from anonymous users who opted to share their data anonymously through a General Data Protection Regulation (GDPR) and California Consumer Privacy Act (CCPA) compliant framework. All final outputs provided to partners are aggregated in order to preserve privacy. The aggregation procedure is specified as data upleveling where some proportions of real locations are deterministically shuffled within Census Block Group (CBG) in the US or geohash level 6 in other countries. This protocol aims to mitigate the risk of re-indetification without affecting the analysis in this study since we infer socioeconomic status at a level broader spatial delineation namely census tract as we discuss further in details in the following section. 

In the actually analysed dataset, the starting point for all observed city is January 2021. Bogota retains longest temporal observation until May 2021, followed by London (February 2021), Jakarta (December 2020), and New York (July 2020). Each individual in every city has a set of trajectories constituting timestamps (start and end) whenever detected at a certain location (latitude and longitude). We focus on mobility traces of people whose home locations are successfully identified at the census tract level as discussed in details in Materials and Methods (Section ~\ref{subsec:inference}). In Bogota, there are approximately 55,000 people containing 25 million trajectories. The number of people fluctuates among cities, so do total trajectories: Jakarta (around 65,000 people/26 million trajectories), London (almost 200,000 people/ 115 million trajectories), and New York (about 277,000 people/30 million  trajectories). To check the general reproducibility of mobility pattern in New York, we also use the SafeGraph dataset~\cite{kang2020multiscale}, which is available at coarser resolution (census tract level) and longer temporal coverage (until May 2021) which is presented in SM Section F. 


\begin{center}
\centering
\resizebox{7.5 cm}{!}{\begin{tabular}{lrrr}
  \hline
Urban Area & Number of People & Number of Trajectory \\ 
  \hline
 Bogota  & 55,000 & 25 million\\
 \hline
 Jakarta & 65,000 & 26 million\\
 \hline
 London  & 200,000  &  115 million\\
  \hline
 New York &  277,000 & 30 million\\
 \hline
\end{tabular}}
\captionsetup{width=0.8\linewidth}
\captionof{table}{\footnotesize\textbf{Sample size}. We have different size sample across cities but preserves the temporal representation of pandemic cycle: before lockdown, lockdown, and reopening.}
\label{tab:1}
\end{center} 

We overlay socioeconomic layer on top of the existing mobility layer. Income related features are fitted for this purpose. In Bogota, multidimensional poverty index ~\cite{mpi2018} at urban section developed by Colombian Bureau of Statistics (DANE) becomes the basis for socioeconomic status computation. It captures quite comprehensive dimension of individual well-being: health, education, utilities and housing, as well as employment. A simpler version of poverty index called poverty rate ~\cite{bps2018} is used in Jakarta at village-level resolution, taking the proportion of people living below particular amount of average monthly income. Meanwhile, socioeconomic configuration of London and New York is plotted respectively based on total annual income recorded by Office for National Statistics (ONS) ~\cite{ons2015} in 2015 at middle layer super output area (MSOA) level and per capita income in 2018 at census tract level taken from American Community Survey (ACS) ~\cite{acs2018}. In each city, we group the people by income distribution in the dataset into 10 equally populated groups from the lowest SES/poorest (1) to highest SES/riches (10). It should be taken into account that direct comparison between cities could not fully established because of diverse characterisation by nonidentical SES indicators and different spatial resolution they are provided at. Nevertheless, comparison across period of the same city is possible to derive in this context.

To synchronise the movement along mobility points and to derive observable structural break in mobility pattern induced by the epidemiological outbreak and policies coming after, we refer to the stringency index on Oxford COVID-19 Government Response Tracker (OxCGRT) dataset ~\cite{hale2021global}. We validate this with actual implementation at city level to ensure policy alignment between national and local government.

\subsection{Algorithm pipeline and inference}
\label{subsec:inference}
We construct an algorithm to detect home and POI (non-home) locations. Our methodology combines the spatial and temporal attributes such as frequency of visit, time window of visit, as well as duration of stay at given locations. We take a further step to infer socioeconomic status for each people (based on home location and POI) by performing spatial projection and merge it with demographic data (average income) from bureau of statistics.

\begin{figure*}[ht!]
    \centering
        \includegraphics[width=0.8\linewidth]{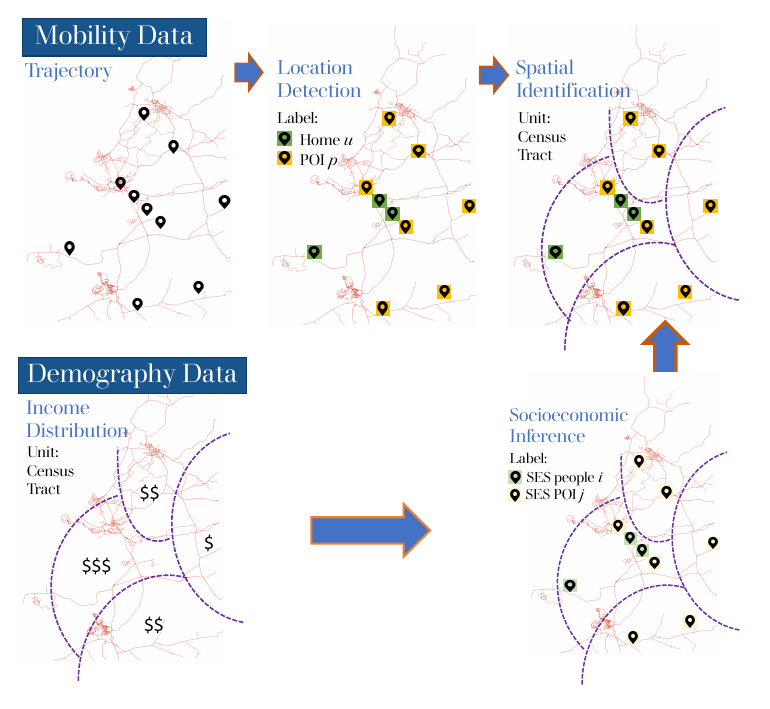}
        \caption{\small{\textbf{Inference Algorithm.} Mobility data contains information regarding whereabouts of people namely geographic locations and timestamp (trajectory). Demographic data covers average income of given spatial unit (eg: census tract). We build an algorithm to separate home $u$ and POI locations $p$ and identify the inferred income based on its spatial delineation. Discretisation on distribution of inferred income results in two separated SES label: SES People $i$ and SES POI $j$.}}
        \label{fig:5}
\end{figure*}

\emph{Home Location:} Detecting home location is a primary step in dealing with mobility data because spatial identifier serves as an intermediary information that allows to couple heterogeneous source of data, including census data. Various decision rules have been developed to identify the whereabouts of people reside. In mobility literature,  a single rule home detection algorithm is widely applied in both continuous (e.g.: global positioning system/GPS data) and non-continuous location traces (e.g.: call detailed record/CDR data)  ~\cite{calabrese2013understanding, csaji2013exploring, phithakkitnukoon2012socio}. Home is defined as the location where highest proportion of activities occurs during night hours with variations regarding time window. To compensate the unavailability of ground truth to be used as validation set, we design more conservative algorithm in determining home location by combining these criteria: a point where an individual is mostly located between 9PM to 6AM for uninterrupted duration at least 6 hours. It results in 50\% people in our dataset of which home locations are being successfully identified.

\emph{POI Location:} Apart from home, human individual activities evolve around other areas for some reasons, including work. Trip between home and work location dominates daily mobility, while visits to other locations are broadly distributed with short inter-event times ~\cite{schneider2013unravelling}. We set criteria for POI location as place other than home where people with identified home locations are present during weekdays from 9AM until 3PM. Afterwards, the rest of locations that do not fall into either home or work category are labelled as others. 

\emph{Socioeconomic Status (SES):} We assign SES label to every individual and and POI based on socioeconomic data. The first step to SES people is to identify socioeconomic feature of area where they live (home location). Similarly, SES POI is inferred by mapping out the area where points (work and other locations) are spatially positioned. We sorted the values by ascending order and split them into equally populated bins of 10 SES labels, making SES 1 to be the poorest and SES 10 to be the richest.

\subsection{Mobility matrix}
\label{subsec:matrix}
In Section~\ref{sec:res}, we rely on the basic formulation of stratification extracted from the \emph{mobility stratification matrix} $M_{i,j}$ that is defined based on the \emph{mobility network} $G=(U,P,E)$. The network $G$ is a bipartite graph that connects people $u$ in the set of node $u\in U$ and POI $p$ from set of node $p\in P$ if $u$ visited $p$, represented as a link $e_{u,p}\in E$ exists. Frequency of visit is counted as edge weights $w_{u,p}$. Stratification is introduced in the network by labelling class membership $c_u=i\in C_U$ to every people and  $c_p=j\in C_P$ to every POI based on their inferred income. As defined earlier in ~\cite{hilman2022socioeconomic}, we have:

\begin{equation}
M_{i,j}= \frac{\sum_{U,c_u=i}\sum_{P,c_p=j} w_{u,p}}{\sum_{j\in C_P}\sum_{U,c_u=i}\sum_{P,c_p=j} w_{u,p}},
\end{equation}

\noindent where the probability of frequency of visits (matrix elements $a_{ij}$) is generated by column-wise normalisation
(SES People $i$) of the frequency matrix. As an example for a mobility stratification matrix see Fig.~\ref{fig:1}.

Given a pair of \emph{mobility stratification matrix} $M_{i,j}$ in two consecutive periods, we define \emph{mobility adjustment matrix} $S_{i,j}$ where the matrix element $b_{ij}$ entails the difference in proportion of frequency of visits. More formally:

\begin{equation}
S_{i,j}= M_{i,j}^{t_1} - M_{i,j}^{t_2},
\end{equation}

\noindent where $t_1$ denotes the initial period and $t_2$ is the succeeding rolling period. For instance, if we have three periods namely Before Lockdown (BL), Lockdown (L1) and Reopening (R1), we could generate three $S_{i,j}$ respectively:

\begin{equation}
S_{i,j}^{BL-L1}= M_{i,j}^{BL} - M_{i,j}^{L1},
\end{equation}

\begin{equation}
S_{i,j}^{L1-R1}= M_{i,j}^{L1} - M_{i,j}^{R1},
\end{equation}

\begin{equation}
S_{i,j}^{BL-R1}= M_{i,j}^{BL} - M_{i,j}^{R1},
\end{equation}

\noindent while $S_{i,j}^{BL-R1}$ shows the difference between period before enforcement of lockdown and reopening (removal some mobility restrictions in the post-lockdown). The result of this computation is provided in Fig.~\ref{fig:4}.

The degree of socioeconomic isolation is computed by the assortativity of the mobility stratification matrix. This \emph{mobility assortativity coefficient} $r$ ~\cite{newman2003mixing, dong2020segregated, bokanyi2021universal} is computed based on the Pearson correlation between row $i\in c_u$ and column $j\in c_p$

\begin{equation}
\mbox{\small\( %
\ r_N = \frac{\sum_{i,j} i j N_{i,j} - \sum_{i,j}i N_{i,j}\sum_{i,j} j N_{i,j}}{\sqrt{\sum_{i,j} i^2 N_{i,j} - \left(\sum_{i,j} i N_{i,j}\right)^2}{\sqrt{\sum_{i,j} j^2 N_{i,j} - \left(\sum_{i,j} j N_{i,j}\right)^2}}}. \)}%
\end{equation}

Values closer to 1 indicate the higher concentration of visiting venues within own socioeconomic range, while lower cutoff values at -1 reveals the tendency of visiting places outside own class. If the value is equal to 0, this measure indicates dispersion in visiting pattern throughout classes without any structural choice preference regarding socioeconomic status of places.

\subsection{Mobility entropy}
\label{subsec:entropy}
Mobility entropy is measured on the basis of generic Shannon's formula \cite{shannon2001mathematical}. In the context of mobility, entropy could be employed to quantify predictability of a visiting pattern. Generally, higher entropy is in line with lower predictability, eliciting the more heterogeneous preference of places to visit in all individual trajectory. At first, we define (\emph{spatial mobility entropy} ${H_m(X)}$) where m is a notation for spatial mobility at individual level in Fig.~\ref{fig:5}a as:

\begin{equation}
H_m{(X)} = - \sum_{x \in X} p_{(x)} \log_2 p_{(x)} = E [- \log p_{(X)}]
\end{equation}

\noindent where $x$ is a discrete random variable representing geographic from all possible location in X of POI locations visited by people.  

We replicate above formulation to measure (\emph{ses mobility entropy} ${H_s(X)}$) in Fig.~\ref{fig:5}b such that:

\begin{equation}
H_s{(X)} = - \sum_{x \in X} p_{(x)} \log_2 p_{(x)} = E [- \log p_{(X)}]
\end{equation}

\noindent where $x$ is replaced by a discrete random variable representing the SES of POI where an user visited.

The value is normalised for each period, therefore the maximum value $1$ and minimum value $0$ is comparable across temporal snapshots. Upper bound value ${H_m(X)} = 1$ implies the sporadic visit to heterogeneous POI locations, while lower bound value ${H_m(X)} = 0$ indicates homogeneous visit pattern to rather limited POI locations. In parallel, ${H_s(X)} = 1$ (heterogeneous SES POI) shows visit to places located in various socioeconomic classes and ${H_s(X)} = 0$ signifies visit pattern characterised by strictly preferred socioeconomic class (homogeneous SES POI).

\subsection{Restriction impact}
\label{subsec:ranking}
We aims to identify the kind of restriction that significantly contributes to changes of diversity in visiting pattern and quantify the magnitude brought by those interventions. To rule out the effectiveness of each type of restrictions, we initiate multivariate linear regression model. There are $k=[1, ..., 9]$ restrictions listed as NPI respectively closings of schools and universities (C1),  closings of workplaces (C2), cancelling public events (C3), limits on gatherings (C4), closing of public transport (C5), orders to stay-at-home (C6), restrictions on movement between cities/regions (C7),  restrictions on international travel(C8) and presence of public information campaigns (H1). Stringency value $S$ for every restriction in each temporal snapshots is obtained from OxCGRT dataset and to be used as independent variable. The dependent variable is two types of mobility entropy, being computed separately: geographic space-based $H_m{(X)}$ and socioeconomic space-based $H_s{(X)}$. 

To further understand the impact magnitude of a single restriction $k \in K$ at timestamp $t \in T$, we fit the data to this form:
\begin{equation}
H_m{(X)}^t \sim   \{ S_{k}^t \}
\end{equation}
and 
\begin{equation}
H_s{(X)}^t \sim   \{ S_{k}^t \}.
\end{equation}
In the equation above, $\{ S_{k}^t \}$ denotes  a set of variables that represents each type of mobility restriction in NPI. The regression covariates indicate the magnitude of restriction impact on segregation. In details, negative values of those covariates imply reduction in the degree of individual spatial and socioeconomic exploration due to respected mobility restrictions. Therefore, the ratio between a pair of the restriction coefficients allow us to compare different impact sizes.

\paragraph{Acknowledgement:}
The authors are thankful for CUEBIQ for providing access to the mobility data. MK acknowledges support from the DataRedux ANR project (ANR-19-CE46-0008), the SoBigData++ H2020 project (SoBigData++ H2020-871042), the SAI Horizon 2020/CHIST-ERA project that was supported by FWF (I 5205-N), the EmoMap CIVICA project, and the National Laboratory of Health Security (RRF-2.3.1-21-2022-00006).

\paragraph{Authors contribution:}
RMH, VS, MGH and MK conceived the study. RMH and MK designed methodology and analyzed the data. RMH, VS, MGH and MK wrote the manuscript with input from all co-authors.

\clearpage

\setcounter{section}{0}

\LARGE
\hspace{-.2in}\textbf{Supplementary Materials}

\normalsize

\renewcommand\thesection{\Alph{section}}

\section{Data and Pipeline}
\label{SM A}

\begin{figure*}[ht!]
    \centering
        \includegraphics[width=1\linewidth]{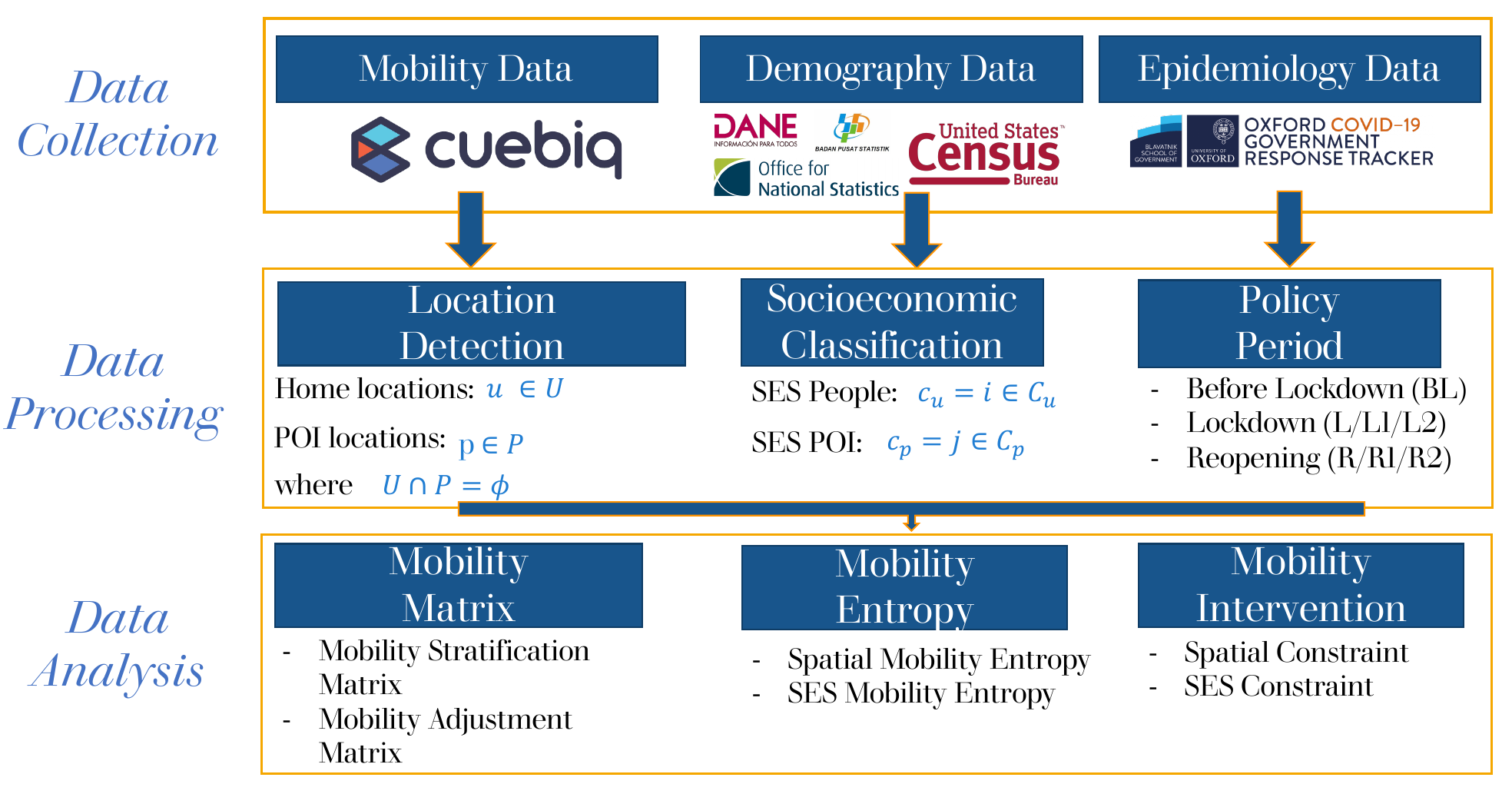}
        \caption{\small{\textbf{Data analytical pipeline.} We observe mobility in Bogota (Colombia), Jakarta (Indonesia), London (United Kingdom), and New York (United States). Three types of data are used: mobility data (Cuebiq), socioeconomic data (Bureau of statistics), and COVID data (OxCGRT/national task force).}}
        \label{SM_A}       
\end{figure*}

Human mobility captures multi-layer information with high spatiotemporal resolution. Not only physical movement from one point to million others, it resumes individual behavioural dynamics in exploring spatial boundaries. In order to make meaningful observation related to individual mobility patterns within urban landscape, we map out socioeconomic condition of people and places they visit by inferring income-based metadata gathered from bureau of statistics of respected locations. This method allows us to comprehensively analyse two aspects of individual trajectory over places: spatial and socioeconomic status (SES) distribution. We construct a pipeline comprising data collection, data processing, and data analysis as depicted in Fig.~\ref{SM_A}. 

The mobility interventions are retrieved from Oxford Covid-19 Government Response Tracker (OxCGRT) dataset, containing nine categories over pandemic period. The list of interventions are provided in Table~\ref{tab:1}. 

\begin{center}
\centering
\resizebox{7.5 cm}{!}{\begin{tabular}{lrrr}
  \hline
Code & Description  \\ 
  \hline
 C1  & School closing \\
 \hline
 C2 & Workplace closing \\
 \hline
 C3  & Cancel public events \\
  \hline
 C4 &  Restrictions on gatherings\\
   \hline
  C5  & Close public transport \\
 \hline
 C6 & Stay at home requirements \\
 \hline
 C7  & Restrictions on internal movement\\
  \hline
 C8 &  International travel controls\\
 \hline
  H1 &  Public information campaigns\\
 \hline
\end{tabular}}
\captionsetup{width=0.8\linewidth}
\captionof{table}{\small{\textbf{Non-pharmaceutical intervention (NPI)}. There are nine restrictions included in this data.}}
\label{tab:1}
\end{center} 

\section{Mobility stratification matrix}
\label{SM B}

\subsection{All visits}
\label{SM B1}
\begin{figure*}[ht!]
    \centering
        \includegraphics[width=0.8\linewidth]{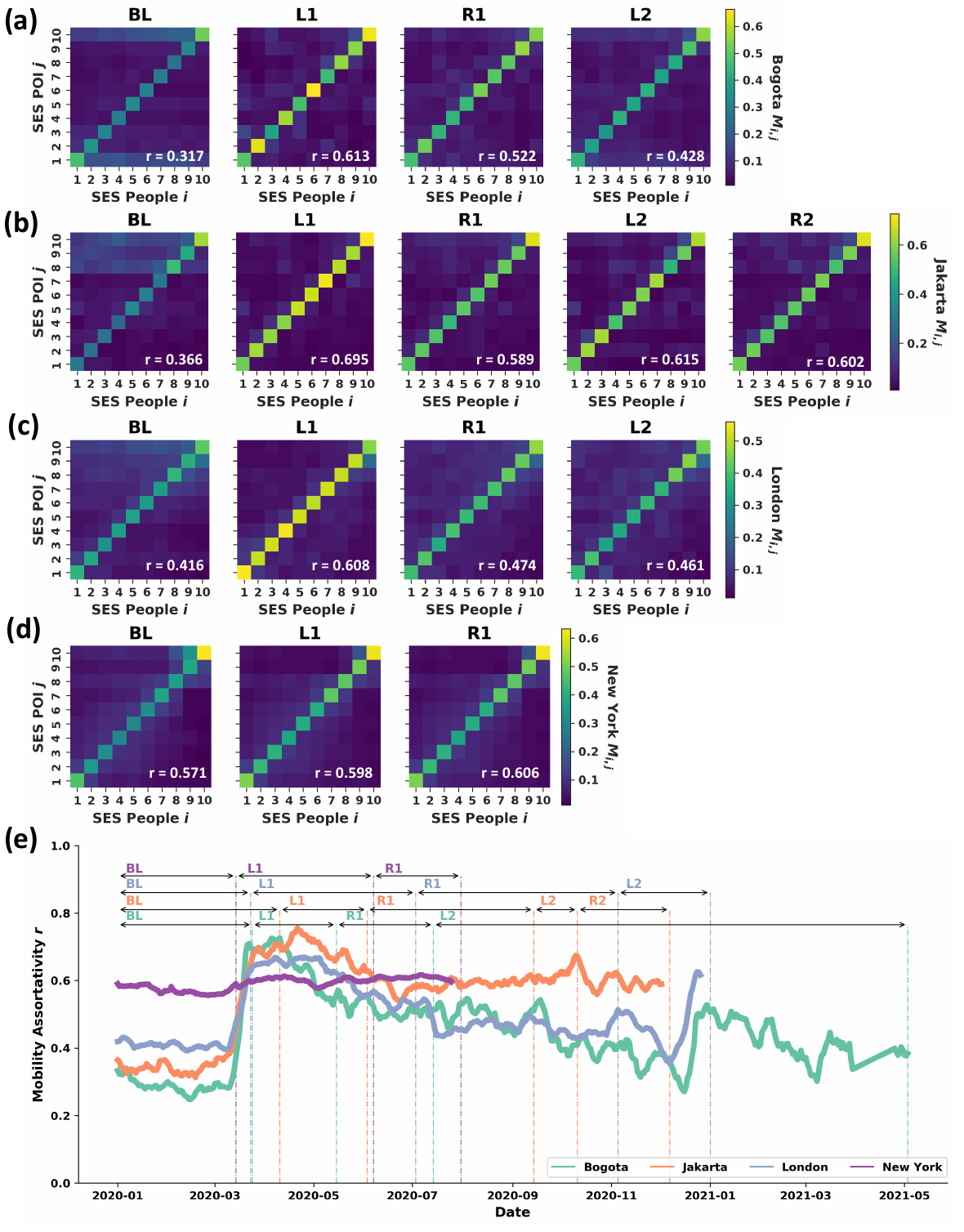}
        \caption{\textbf{Mobility Stratification Matrix for all visits $M_{ij}$.} Matrix elements in Fig.~\ref{fig:SM_B1}a-d represent the magnitude of frequency visits for each pair of SES People $i$ and SES POI $j$ where lighter colour shows larger visit proportion. All locations found in individual trajectories are taken into account.}
        \label{fig:SM_B1}       
\end{figure*}

Distribution of frequency visit with regards to socioeconomic stratification between SES People $i$ and SES POI $j$ is conceptually introduced in Section 5.2 as Mobility Stratification Matrix $M_{ij}$. Normalisation in performed by own SES (column-wise). Fig.~\ref{fig:SM_B1} reveals the generic pattern in which assortative mixing increases during the lockdown as increasing $r$ is found across cities. It reflects the extend individual responds to the pandemics by reorganising their typical mobility configuration. In the case of more than one period of lockdown appears (L1 and L2), the first seems to be stronger in inducing the isolation effect. As the reopening (R1) phase is started, the assortative visit remains higher than the level before lockdown (BL).

\subsection{Without home area visit}
\label{SM B2}

\begin{figure*}[ht!]
    \centering
        \includegraphics[width=0.8\linewidth]{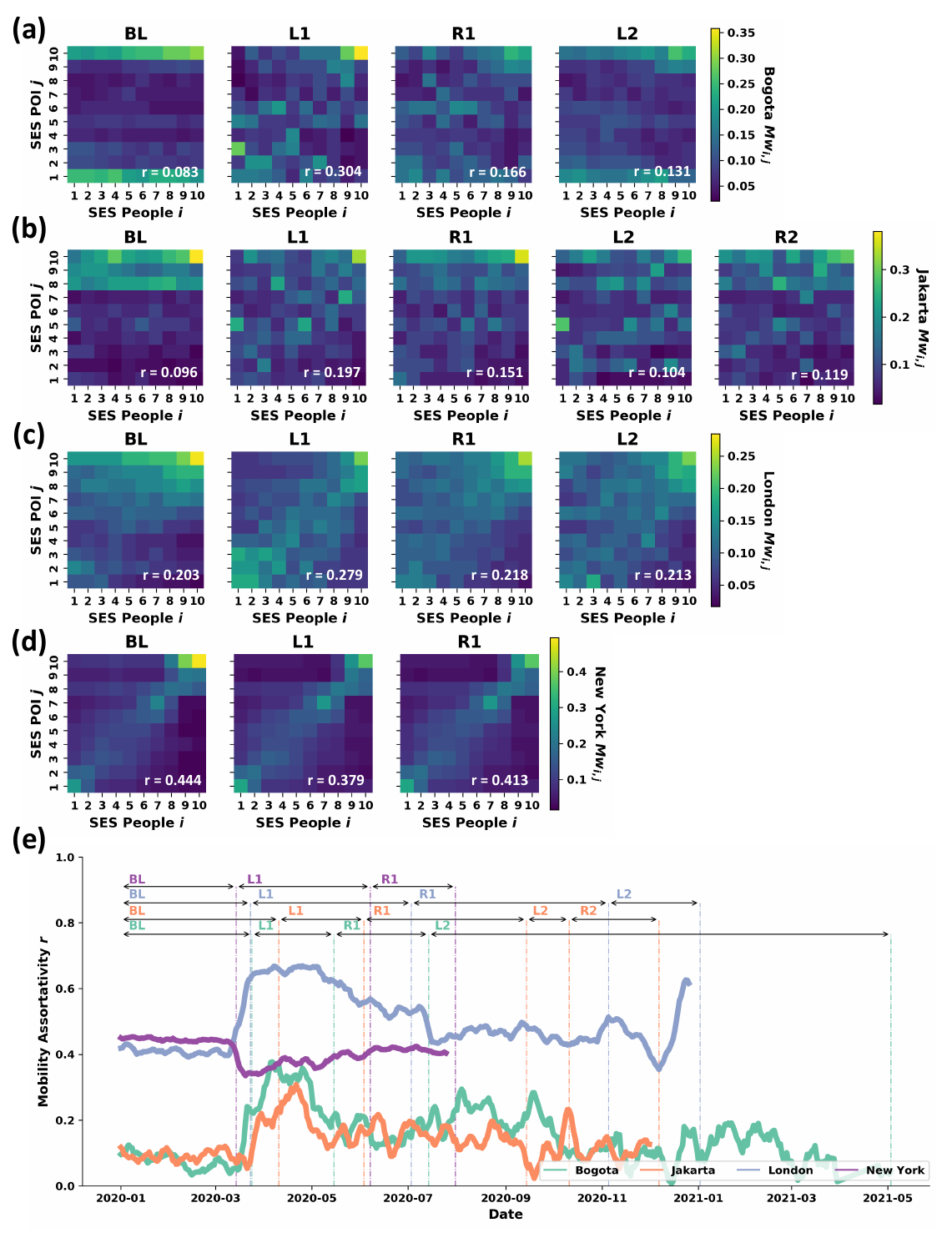}
        \caption{\textbf{Mobility Stratification Matrix for visits outside home area $Mc_{ij}$.} Proportion of frequency visit of people from SES $i$ to places in SES $j$ is computed after removing places located in own neighbourhood. The lighter bin colour, the higher visit probability is.}
        \label{fig:SM_B2}       
\end{figure*}

We repeat the procedure used to generate Fig.~\ref{fig:SM_B1} after excluding local visits to own neighbourhood to generate Mobility Stratification Matrix for visits outside home area $Mc_{ij}$. This step is considered as robustness control over the persistent assortative mixing. In Fig.~\ref{fig:SM_B2} we see that the first lockdown is still the most stringent because it alters preference to visit more places within own socioeconomic class. Comparing to Fig.~\ref{fig:SM_B1}, assortativity coefficient $r$ in general is away lower, indicating that short distance visit in the surrounding neighbourhood assumes considerable proportion on mobility pattern.

\section{Mobility adjustment matrix}
\label{SM C}

\begin{figure*}[ht!]
    \centering
        \includegraphics[width=0.9\linewidth]{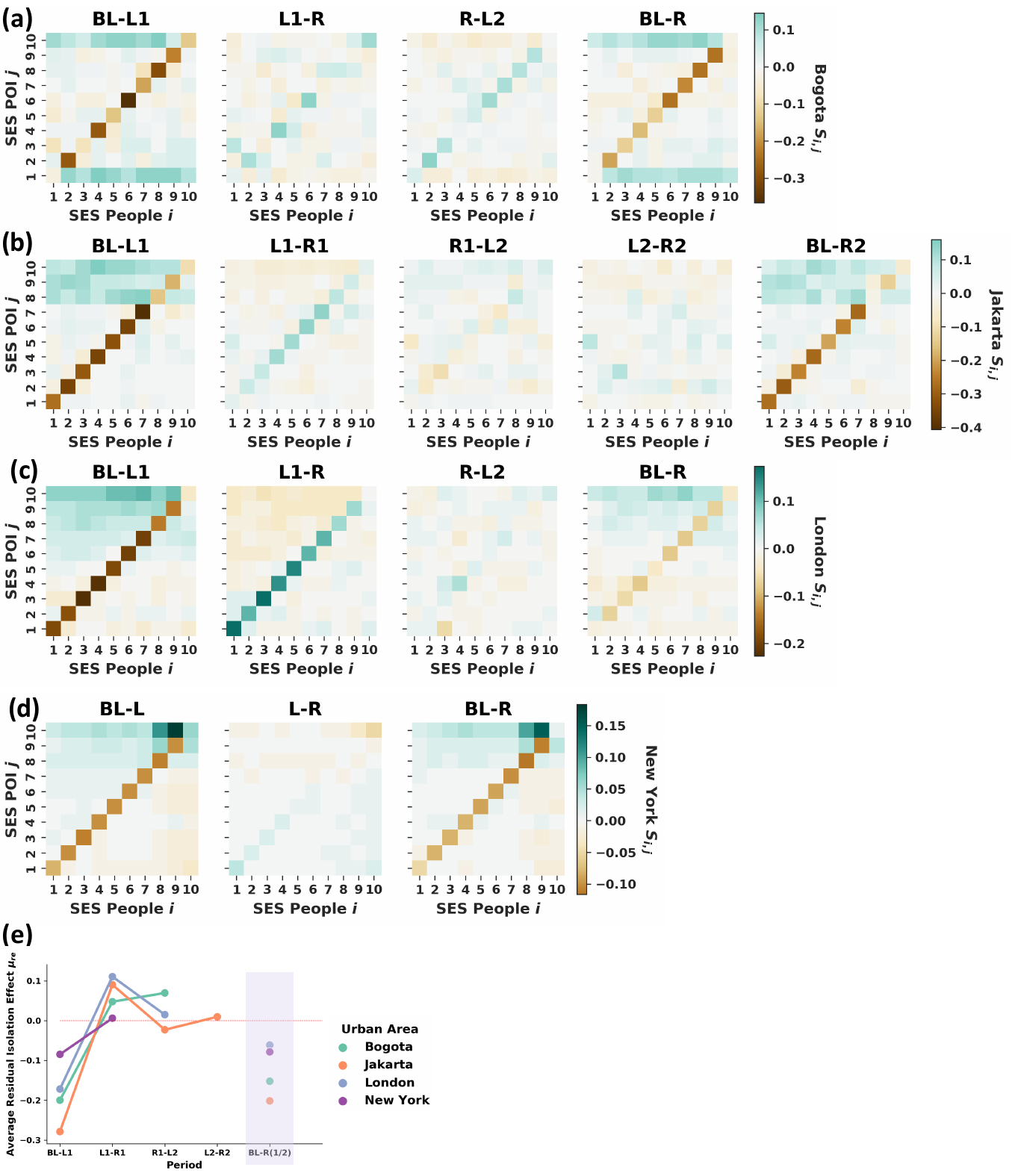}
        \caption{\textbf{Mobility Adjustment Matrix for all visits $S_{ij}$.} The difference in term of visit probability between a pair of two consecutive Mobility Stratification Matrix $M_{ij}$ is measured. The presence of white bins indicates indifferent visiting pattern, while green shows more visits during the first period. Otherwise, brown shades appear. All locations found in individual trajectories are taken into account.}
        \label{fig:SM_C1}       
\end{figure*}

Mobility adjustment matrix $S_{ij}$ is constructed to detect the indication of residual isolation effect. We operationalise the computation in Section 5.2 in which the difference in proportion of frequency visits between two consecutive periods is visible in Fig.~\ref{fig:SM_C1}. None of cities in this study exhibit full recovery after the occurrence of reopening as the bin colour remains under brown shades, indicating larger visit ratio to places in own socioeconomic class as to compare with before lockdown period. Therefore, it leads to the notion of residual isolation induced by COVID outbreak. 

\subsection{All visits}
\label{SM C1}

\subsection{Without home area visit}
\label{SM C2}

Mobility adjustment matrix $S_{ij}$ is transformed to $Sc_{ij}$ by eliminating visits to own neighbourhood. It runs on similar motivation in Section ~\ref{SM C2}, namely as robustness check given the large local visits in the individual trajectory. 

\begin{figure*}[ht!]
    \centering
        \includegraphics[width=0.8\linewidth]{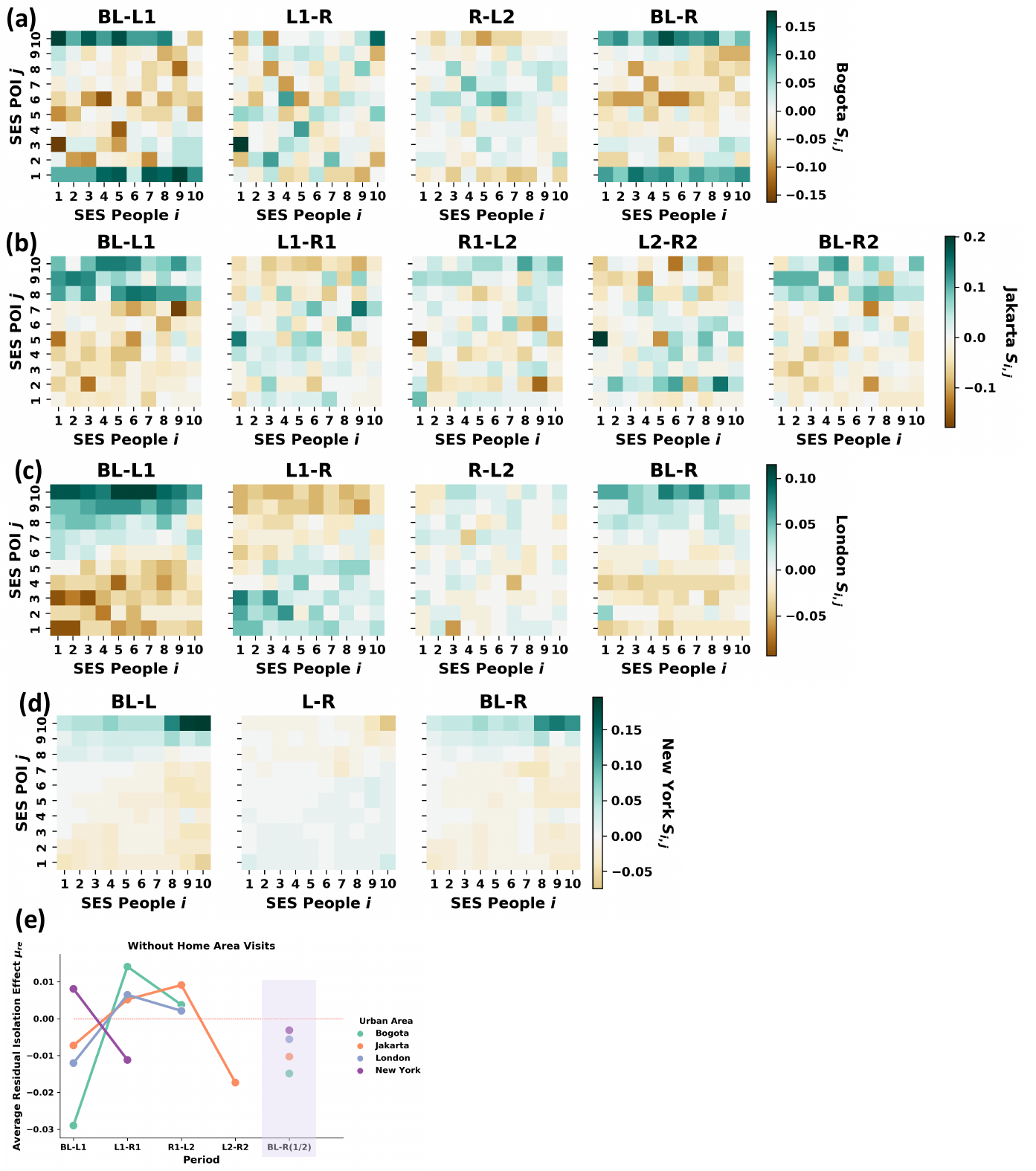}
        \caption{\textbf{Mobility Adjustment Matrix for visits outside home area $Sc_{ij}$.} Every Mobility Stratification Matrix for visits outside home area $Mc_{ij}$ is paired with the one in the following period. There are three patterns to detect: no difference between those two periods (white), dominant visit in the first period (green), and dominant visit in the second period (brown).}
        \label{fig:SM_C2}       
\end{figure*}

Fig.~\ref{fig:SM_C2} tries to uncover the main attribution of residual isolation effect by eliminating visits to own neighbourhood/home area. This procedure dilutes the magnitude of assortativity force, therefore we address the residual isolation effect as a longer term consequence of localised mobility due to COVID restrictions. 

Interestingly, BL-R shows segregated pattern of visit where before lockdown people tend to explore more places in higher socioeconomic ranks (top rows/green shades) while during the reopening places in lower classes contribute more to visit proportion (brown shades) in every cities. Beyond that, Bogota exhibit bimodal segregation where dominant visit before lockdown does not only happen in upper class, but also lower class.  

\section{Mobility entropy}
\label{SM D}

\subsection{Spatial mobility entropy}
\label{SM D1}

Heterogeneity of places visited by individual is quantified by computation of Spatial Mobility Entropy ${H_m(X)}$ proposed in Section 5.3. Dispersion of value may take either to the direction of 0, signifying strict preference on particular locations over the rest and making the trajectory more homogeneous spatial wise. In contrast, as the value takes closer to 1, no strict preference presumed and visits are widely distributed across locational space. We find that people become more restricted in deciding which locations to visit as the average value ${H_m(X)}$ hits the lowest point than ever in all cities. The introduction of reopening phase does not directly bounce the value back to the normal level before lockdown, in line with condition suggested in Fig.~\ref{fig:SM_B1} and Fig.~\ref{fig:SM_C1}. 

\begin{figure*}[ht!]
    \centering
        \includegraphics[width=1\linewidth]{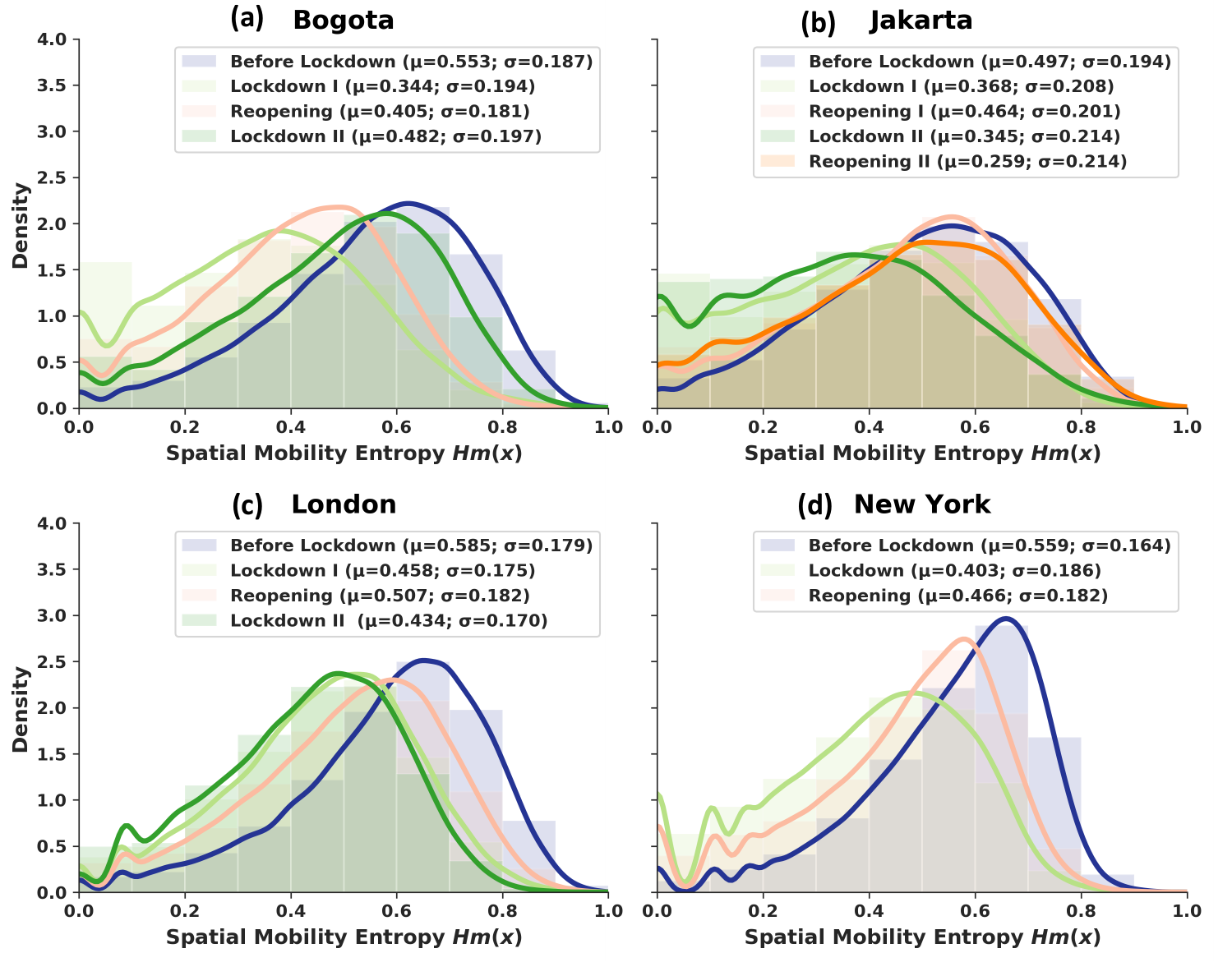}
        \caption{\textbf{Spatial Mobility Entropy ${H_m(X)}$.} We measure heterogeneity of individual preference regarding location of places visited. The presence of commonly repeated places pushes the value closer to zero, denoting lower degree of heterogeneity. On the other hand, higher variability of locations is represented by value near 1.}
        \label{fig:SM_D1}       
\end{figure*}

\subsection{Socioeconomic mobility entropy}
\label{SM D2}

\begin{figure*}[ht!]
    \centering
        \includegraphics[width=1\linewidth]{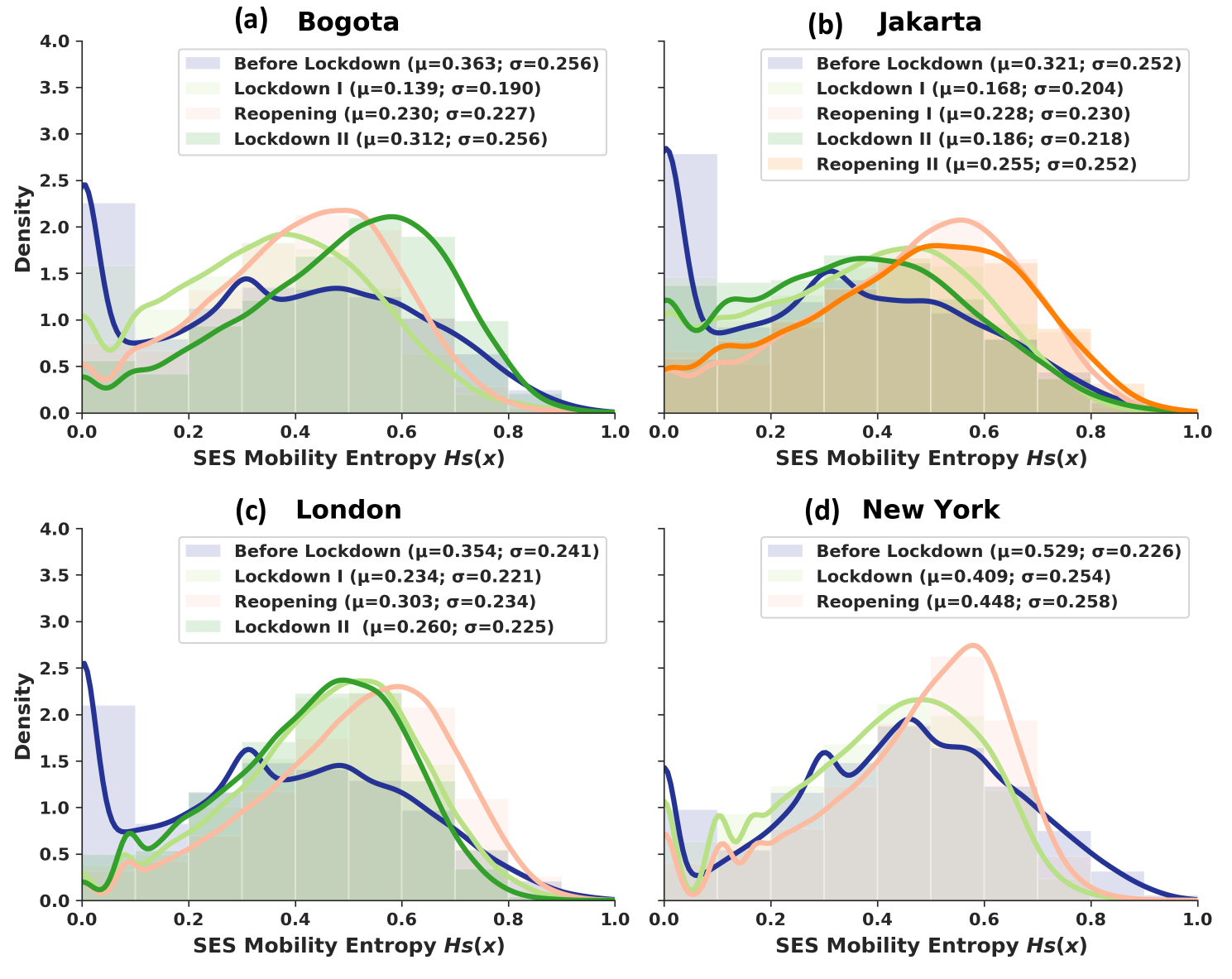}
        \caption{\textbf{Socioeconomic Mobility Entropy ${H_s(X)}$.} After replacing geolocation of places in individual trajectory by SES information, we recompute entropy. As the value skews to 0, visiting pattern tends to be concentrated on particular SES, otherwise it is somewhere close to 1.}
        \label{fig:SM_D2}       
\end{figure*}

In this section, we redo the computation for trajectory heterogeneity in terms of socioeconomic factor based on entropy formulation in Section 5.3. To measure Socioeconomic Mobility Entropy ${H_sX)}$, we substitute geolocation feature with SES of places. The result in Fig.~\ref{fig:SM_D2} confirms previous finding where people have stricter preference over places during lockdown. It is beyond spatial boundary since socioeconomic profile of those places is now also heavily skewed, making average value ${H_s(X)}$ touches lowest record in comparison to other periods. Therefore, it reaffirms condition stipulated in Fig.~\ref{fig:SM_B1}, Fig.~\ref{fig:SM_C1} and Fig.~\ref{fig:SM_D1}. 

\section{Robustness of mobility adjustment}
\label{SM F}

We take into account the robustness check of isolation effect by applying Kruskal-Wallis H Test (non-parametric one-way ANOVA) on Mobility Stratification Matrix for both before ($M_{i,j}$) and after removing visits to own home area ($Mw_{i,j}$). The formulation of the null hypothesis ($H0$) could be defined as an equal median between before lockdown and another period that comes after. If the $p$-value appears to be smaller than the confidence level $\alpha=0.05$, $H0$ is rejected. Otherwise, he alternative hypothesis ($H_a$) remains. Table \ref{tab:F1} and \ref{tab:F2} provide justification for the presence of different degrees of isolation effect due to the variability of mobility in response to the dynamics of mobility restrictions. New York stands on strikingly opposite pattern as statistically significant difference is seen after removing local visits to the area where home is located while other cities exhibit such pattern for broad visits to any locations.

\subsection{All visits}
\label{SM F1}

\begin{table}[ht]
\centering
\resizebox{15 cm}{!}{\begin{tabular}{lrrrrrrrr}
  \hline
Urban Area & Matrix Element & BL \& L1  & L1 \& R1 & R1 \& L2 &  L2 \& R2 &  BL \& R1 &  BL \& R2 \\ 
  \hline
 Bogota  & all  & $7.556^{*}$ & $3.567{*}$ & 1.664 & \textemdash & 0.435 & \textemdash \\
 \hline
  Bogota  & diagonal  & $11.063^{*}$ & 2.063 & $7.406^{*}$ & \textemdash & $9.606^{*}$ & \textemdash \\
 \hline
 Jakarta & all  & $9.135^{*}$ & $5.108^{*}$ & 0.748 & 0.043 & 1.504  & 2.720  \\
 \hline
  Jakarta & diagonal  & $12.091^{*}$ & $10.079^{*}$ & 1.651 & 0.571 & $9.143^{*}$  & $9.606^{*}$  \\
 \hline
 London &  all  & $10.832^{*}$   &  $12.362^{*}$ & 0.215 & \textemdash  & 0.299 & \textemdash  \\
  \hline
  London &  diagonal  & $14.286^{*}$   &  $13.719^{*}$ & 1.286 & \textemdash  & $9.143^{*}$ & \textemdash  \\
  \hline
 New York  &  all &  1.404 & 5.970 & \textemdash & \textemdash & 1.381 & \textemdash    \\
 \hline
  New York  &  diagonal &  $7.406^{*}$ & 0.143 & \textemdash & \textemdash & $6.606^{*}$ & \textemdash    \\
 \hline
\multicolumn{1}{l}{\textsuperscript{*}$p<0.05$}
\end{tabular}}
\caption{\textbf{Kruskal-Wallis H Test on Mobility Stratification Matrix before removing visits to home area across pairs of policy period  ($M_{i,j}$)}. Statistical significance could be implied in which the induced isolation effect largely takes place between before lockdown and the first lockdown (BL \& L1). It happens in all urban areas (for diagonal elements) but New York (for all elements) as the  $p$-value is away lower than the confidence level at $\alpha=0.05$. Even after the introduction of the first reopening, the distribution of mobility pattern still does not revert to the pre-pandemic level (BL \& R1)}
\label{tab:F1}
\end{table}

\subsection{Without home area visit}
\label{SM F2}

\begin{table}[ht]
\centering
\resizebox{15 cm}{!}{\begin{tabular}{lrrrrrrrr}
  \hline
Urban Area & Matrix Element & BL \& L1  & L1 \& R1 & R1 \& L2 &  L2 \& R2 &  BL \& R1 &  BL \& R2 \\ 
  \hline
 Bogota  & all & 1.728 & 0.795 & 0.202 & \textemdash & $6.595^{*}$ & \textemdash \\
 \hline
  Bogota  & diagonal & 1.851 & 0.006 & 0.001 & \textemdash & 1.463 & \textemdash \\
 \hline
 Jakarta & all & $6.090^{*}$ & 0.006 & 0.013 & 0.160 & $4.550^{*}$  & 3.252  \\
 \hline
  Jakarta & diagonal & 1.286 & 0.051 & 0.001 & 0.281 & 0.691  & 0.966  \\
 \hline
 London & all & 0.294   &  0.199 & 0.001 & \textemdash  & 0.638 & \textemdash  \\
  \hline
 London & diagonal & 1.286   &  0.463 & 0.023 & \textemdash  & 1.286 & \textemdash  \\
  \hline
 New York  & all & 0.119 & 0.084 & \textemdash & \textemdash & 0.001 & \textemdash    \\
 \hline
 New York  & diagonal & 0.206 & 0.051 & \textemdash & \textemdash & 0.206 & \textemdash    \\
 \hline
\multicolumn{1}{l}{\textsuperscript{*}$p<0.05$}
\end{tabular}}
\caption{\textbf{Kruskal-Wallis H Test on Mobility Stratification Matrix after removing visits to home area across pairs of policy period ($Mw_{i,j}$)}. Mobility pattern differs significantly between before and during the first lockdown (BL \& L1) in Jakarta (for all elements) but not apparent in other urban areas given the  $p$-value is away lower than the confidence level at $\alpha=0.05$. Similar direction also becomes visible between before and during the first reopening (BL \& R1). Strict isolation along diagonal elements is not found anywhere. Therefore, levelling up the contribution of local visits in the surrounding of home locations to isolation.}
\label{tab:F2}
\end{table}

\section{Manhattan Effect}
\label{SM G}

New York is made up of five boroughs respectively Manhattan, Brooklyn, Queens, Bronx, and Staten Island. Among others, Manhattan is the centre of human activity agglomeration. Manhattan as a borough with the highest economic pull-factors in New York is massively affected, because mobility disruption hit not only movement of people inside borough, but also inter-borough movement that usually found in commuting pattern to workplace. People who reside in Brooklyn and Queens, for example, stop commuting to Manhattan as many of them switched to working from home practice. It is also reflected in lower use of public transportation and level of road traffic. 

Segregation pattern changes as a response to mobility restriction imposed due to the pandemic. In Section 2.2, we see that the mobility assortativity $r$ in New York is relatively flat as to compare to other cities such as Bogota, Jakarta, and London, but a more substantial mechanism at work that shapes urban human dynamics might contribute as well. In this section we take two strategies to disentangle spatial scale. At first, we focus in the area of Manhattan where activities and mobilities are heavily concentrated. Later on, we analyse mobilities in each borough that together unite as New York (intra-mobility), followed by mobilities between a pair of boroughs (inter-mobility).

\begin{figure*}[ht!]
    \centering
        \includegraphics[width=1\linewidth]{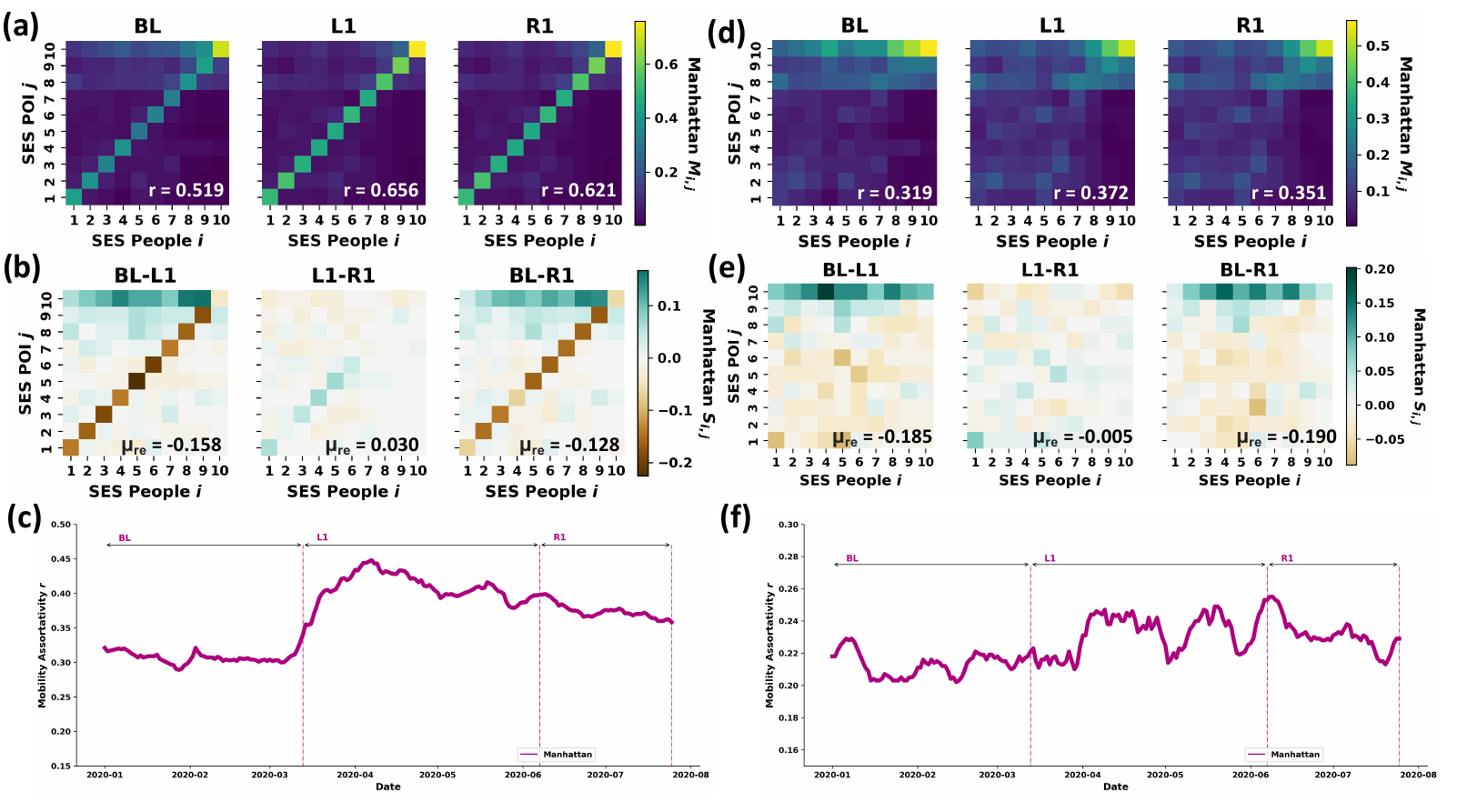}
        \caption{\footnotesize\textbf{Mobility stratification matrix $M_{i,j}$, mobility adjustment matrix $S_{i,j}$, and mobility assortativity $r$}. We impose additional layer of filtering in New York by only looking at the locations within Manhattan boundary. On the left (Fig.~\ref{fig:SM_G1}a-c), we take into account all visits, while on the right (Fig.~\ref{fig:SM_G1}d-f), we remove local visits to home area. Assortative mixing touches the highest level during lockdown ($r=0.656$). After reopening, average residual isolation effect $\mu_{re}$ is still 12.8\% higher as to compare to before lockdown period.}
        \label{fig:SM_G1}       
\end{figure*}

Mobility stratification in Manhattan is visualised as matrix in Fig. ~\ref{fig:SM_G1}a. Homophilic mobility defined as movement within own socioeconomic class during the lockdown is 26\% higher than before lockdown. Emergence of reopening phase does not directly brings back the normal condition since it still exceeds the original level by 20\%. Even after removing local visits (Fig. ~\ref{fig:SM_G1}d), the pattern stands still. This finding is consistent with global pattern previously captured in other cities in this study such as Bogota, Jakarta, and London (see Section ~\ref{SM B1}). 

Taking a pair of matrices in two consecutive periods, we have another form of matrix to show mobility adjustment as seen in Section ~\ref{fig:SM_G1}b. Measures taken during lockdown affect individual preference regarding their mobility. There is increase in visits to places within own socioeconomic range by at least 15\% (see left matrix). Reopening happen at some points, however nothing such fully recovery exists. We still find that the average value of diagonal elements is 12\% higher than before lockdown (see BL-R1). In the case of disregarding dominant local visits to own neighbourhood (Fig. ~\ref{fig:SM_G1}e), average residual isolation effect $\mu_{re}$ in the reopening still surpasses the baseline period before lockdown by 19\%. After all, residual isolation effect remains prominent in Manhattan. 

Sliding window algorithm is implemented to generate Fig.~\ref{fig:SM_G1}c and Fig.~\ref{fig:SM_G1}f. For every 1 week window with 1 day slide
interval, a mobility matrix is generated with computed mobility assortativity $r$.  For both all visits (Fig.~\ref{fig:SM_G1}c) and visits to places other than own neighbourhood (Fig.~\ref{fig:SM_G1}f), increasing $r$ overlaps with lockdown period.

\begin{figure*}[ht!]
    \centering
        \includegraphics[width=1\linewidth]{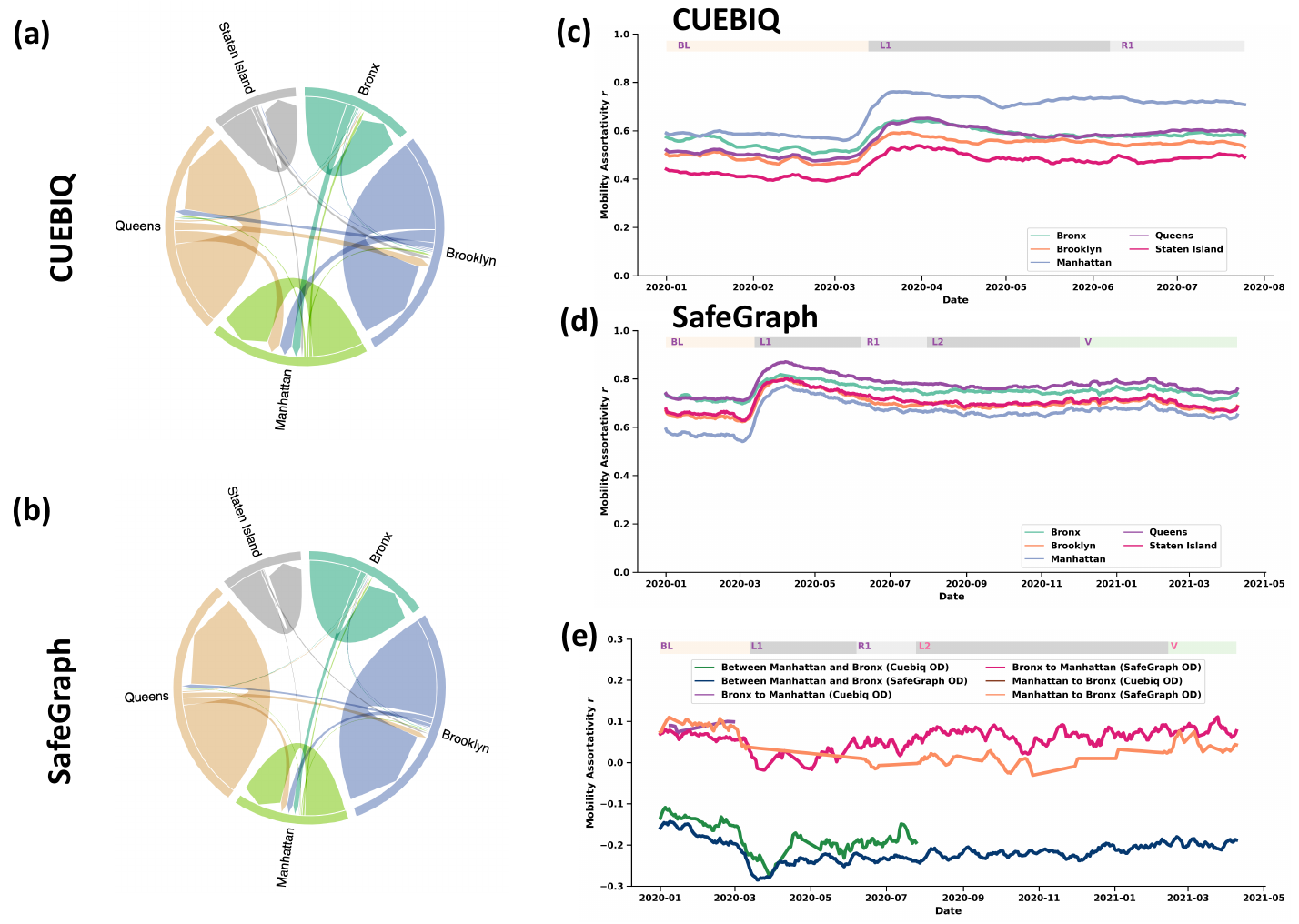} \caption{\footnotesize\textbf{Trip composition and mobility assortativity $r$ by category}. Intra-mobility (mobility within borough) dominates trip proportion in both Cuebiq (Fig.~\ref{fig:SM_G2}a) and SafeGraph dataset (Fig.~\ref{fig:SM_G2}b). Mobility assortativity is computed at census tract level based on OD matrix, showing similar pattern for intra-mobility mixing namely increasing segregation in the two datasets (Fig.~\ref{fig:SM_G2}c-d). Interestingly, segregation in inter-mobility (mobility between borough) tends to be lower instead, for instance in mobility flow between Manhattan and Bronx (Fig.~\ref{fig:SM_G2}e).}
        \label{fig:SM_G2}       
\end{figure*}

Computations for mobility in New York based on Cuebiq dataset (Fig.~\ref{fig:SM_G2}a) are reproduced for SafeGraph dataset (Fig.~\ref{fig:SM_G2}b). The two comes in conformity in terms of the proportion of mobility category in which individual flows within a single borough (intra-mobility) surpasses the fluxes across different territories (inter-mobility). The first is presented in Fig.~\ref{fig:SM_G2}c (Cuebiq) and Fig.~\ref{fig:SM_G2}d (SafeGraph). A striking mirroring degree of assortativity in mobility $r$ within Manhattan is seen, ranging from 0.6 before the implementation of lockdown to 0.8 in the aftermath. While the value of $r$ is slightly different in Bronx (light green), Brooklyn (orange), Queens (purple), and Staten Island (pink), the pattern stays the same: increasing segregation since the lockdown period. One reason behind is that once people stay at residential area, they are bounded not only by spatial scale, but also socioeconomic homogeneity in the surrounding neighbourhoods. 

On contrary, individual flows across boroughs (inter-mobility) exhibits decreasing segregation as shown in Fig.~\ref{fig:SM_G2}e in the case of mobility flux between Manhattan and Bronx. As a undirected mobility network, mobility recorded in Cuebiq dataset (dark green) and SafeGraph dataset (dark blue) indicate the emergence of disassortative mixing with value lower than 0, implying that people abruptly visit places differ from own socioeconomic status whenever they need to step out territory/borough where they reside due to multiple mobility reasons (e.g.: work or school).


\end{document}